\documentclass[aps,prd,twocolumn,superscriptaddress,nofootinbib,longbibliography]{revtex4-2}

\usepackage{amsmath,amssymb,amsfonts,amsthm}
\usepackage{mathrsfs}
\usepackage{bm}
\usepackage{braket}
\usepackage{graphicx}
\usepackage{dcolumn}
\usepackage{color}
\usepackage{xcolor}
\usepackage{hyperref}
\usepackage{physics}
\usepackage{mathtools}
\usepackage{array}
\usepackage{booktabs}
\usepackage{multirow}
\usepackage{enumitem}
\usepackage{pstricks}
\usepackage{orcidlink}
\hypersetup{
    colorlinks=true,
    linkcolor=blue,
    citecolor=blue,
    urlcolor=blue
}

\begin{document}

\title{Operational Limits of Gravity-Mediated Entanglement}

\author{Atta ur Rahman~\!\!\orcidlink{0000-0001-7058-5671}}
\affiliation{School of Physical Sciences, University of Chinese Academy of Sciences, Yuquan Road 19A, Beijing 100049, China}

\author{Muhammad Noman~\!\!\orcidlink{0009-0003-6136-0114}}
\affiliation{Institute of Fundamental Physics and Quantum Technology and School of Physical Science and Technology, Ningbo University, Ningbo 315211, China}

\author{Ying-Jian Zhu~\!\!\orcidlink{0009-0005-7407-2095}} 
\affiliation{Department of Physics, School of Science, Beihua University, 15, Jilin Str. Jilin, 132013, China}

\author{Jia-Xin Peng~\!\!\orcidlink{0000-0002-0119-0139}}
\affiliation{School of Physical Science and Technology, Nantong University, Nantong 226019, China}

\author{Saeed Haddadi~\!\!\orcidlink{0000-0002-1596-0763}}\email{haddadi@ipm.ir}
\affiliation{School of Particles and Accelerators, Institute for Research in Fundamental Sciences (IPM), P.O. Box 19395-5531, Tehran, Iran}

\date{\today}

\begin{abstract}
Tests of gravity-mediated entanglement aim to determine whether two massive objects in spatial superposition can become entangled through their mutual gravitational interaction. In an ideal model, the two objects are often treated as pointlike branches, but real wave packets have a finite width and are also exposed to dephasing and residual non-gravitational forces. We study these effects in a two-branch gravitational cat-state model with Gaussian wave packets. The Newtonian interaction is averaged over the finite wave-packet width, and spatially correlated dephasing is included phenomenologically. We then ask a practical question: when does the generated state remain useful rather than merely entangled? To answer this, we examine Bell violation, teleportation fidelity, coherent information, Fisher information, and state distinguishability. We find finite parameter windows where the same state is Bell-nonlocal and teleportation-useful, but these windows shrink as the wave packets become broader or the dephasing becomes stronger. We also show that a residual electrostatic branch interaction can coherently cancel the gravitational nonlocal phase and strongly suppress the operational signal. The results identify finite-width and electrostatic tolerance boundaries that should be considered in future tests of gravity-mediated entanglement. 
\end{abstract}


\maketitle

\section{Introduction}

A direct test of quantum aspects of gravity does not have to begin at the Planck scale. A more accessible route is to use low-energy massive systems whose center-of-mass states are placed in spatial superposition. If two such systems interact only through their mutual gravitational field, and if the branch-dependent Newtonian phases generate entanglement between them, then the interaction cannot be described as a purely local classical channel within the assumptions of the protocol. This is the central idea behind the Bose--Marletto--Vedral (BMV) class of proposals~\cite{Bose2017,MarlettoVedral2017}. The appeal of this setting is that it connects gravity with experimentally meaningful quantum-information tests. Bell-type criteria ask whether the produced correlations go beyond local realistic descriptions~\cite{Horodecki1995}, while teleportation and coherent-information tests ask whether the same state can act as a useful communication resource~\cite{Bennett1993,Horodecki1999,Devetak2005}. In this way, a weak gravitational phase is not treated only as a small energy shift; it is tested through the operational tasks it can support.

The interpretation of gravity-mediated entanglement is, however, subtle. Entanglement generation between two masses would be a strong indication against a strictly local classical mediator in the usual BMV setting, but it would not automatically settle every possible question about semiclassical, hybrid, or postquantum descriptions of gravity. For this reason, several works have examined what such experiments can and cannot certify. Classical-channel and hybrid-dynamics models have clarified the assumptions under which entanglement witnesses become meaningful~\cite{HallReginatto2018,Belenchia2018}. Relativistic and field-theoretic analyses have also shown that the claim must be phrased carefully, especially when one asks what is meant by the gravitational mediator and which degrees of freedom are being tested~\cite{ChristodoulouRovelli2019,Danielson2022}. More recent studies have further discussed measurement choices, hidden-variable descriptions, and the distinction between observing entanglement and excluding broader classes of classical explanations~\cite{Carney2022,Galley2022,Galley2023}. The present work follows this cautious viewpoint. We do not claim that one witness proves a complete theory of quantum gravity. Instead, we ask which operational signatures remain available once finite wave-packet width, dephasing, and residual force systematics are included~\cite{Christodoulou2023,MartinMartinez2023}.

The experimental motivation for this question has become stronger in recent years. Levitated optomechanics has made the preparation and control of massive quantum motion a realistic research direction~\cite{RomeroIsart2011,Delic2020}. Feedback cooling, optical trapping, and nanoparticle-control techniques have improved the ability to approach the motional regimes needed for massive-superposition experiments~\cite{Magrini2021,Tebbenjohanns2021}. At the same time, matter-wave interferometry with increasingly large particles has pushed the boundary of spatial coherence for massive objects~\cite{Fein2019,Rudolph2022}. Spin-dependent interferometry and related proposals have then translated the BMV idea into more concrete requirements on mass, branch separation, coherence time, shielding, and readout~\cite{Pino2018,Pedernales2020,Neumeier2024}. These developments make the idealized point-particle picture less sufficient. In a real experiment, the branch wave packets have finite width, their coherence is exposed to environmental dephasing, and residual electromagnetic forces may act on the same branch geometry as gravity~\cite{GonzalezBallestero2021,RodaLlordes2024}.

A related literature has studied gravitational cat states and gravity-mediated systems using the language of quantum resources. Anastopoulos and Hu discussed massive systems in spatial superposition in connection with Newtonian-field fluctuations and two-state gravitational behavior~\cite{AnastopoulosHu2015,AnastopoulosHu2020}. Later works investigated how thermal effects and environmental channels affect the quantum correlations of two gravitational cat (gravcat) systems~\cite{RojasLobo2023,Haddadi2024Dense,Haddadi2024EPJC}. These studies are valuable because they show that gravitationally induced correlations can survive under nonideal conditions and that the behavior of a gravcat system can be organized through fundamental concepts such as entanglement, quantum coherence, steering, Bell nonlocality, dense coding capacity, or quantum fidelity~\cite{Rahman2025PRD,Jaloum2025}. However, much of this literature is still centered on whether nonclassical correlations are present. That is an important first question, but it is not the whole story. A state may remain entangled while failing to violate a Bell inequality, failing to beat the classical teleportation threshold, carrying no positive coherent information, or becoming difficult to distinguish from a calibrated non-gravitational mimic~\cite{Wootters1998,BennettWiesner1992,Audenaert2007}. In addition, finite wave-packet width and coherent electrostatic contamination are often treated as secondary corrections, although they can directly change the branch phase that the experiment is trying to observe.

This is the gap addressed in the present work. We treat the finite-width gravitational cat state as a noisy branch channel and ask when it remains useful and distinguishable from the benchmark models considered here. The finite width is included directly through a Gaussian-smeared Newtonian kernel, rather than being added later as a qualitative correction. Spatially correlated dephasing is included through a branch-decoherence matrix, so that the loss of coherence depends on how different the two branch configurations are in space. Starting from the same coherent branch input, we then evaluate Bell--CHSH (Clauser–Horne–Shimony–Holt) violation, optimal teleportation fidelity, coherent information, concurrence, Fisher information, trace-distance discrimination, and Chernoff-copy costs. These diagnostics are not introduced to overstate the result. They are used to separate ordinary entanglement survival from stricter operational usefulness~\cite{Lloyd1997,Devetak2005,Giovannetti2011,Degen2017}.

The most important systematic effect considered here is residual electrostatic contamination. In massive-superposition experiments, electrostatic forces are not merely an additional source of random noise. A residual charge, patch potential, induced dipole, or imperfectly screened Coulomb-like interaction can contribute a coherent branch-dependent phase. If this phase enters the same local-phase-invariant branch combination as the gravitational phase, it can reduce or even cancel the nonlocal phase that supports Bell violation, teleportation advantage, metrological sensitivity, and discrimination performance. This cancellation does not remove gravity from the Hamiltonian. It means that the operational signal associated with the gravitational branch phase can be suppressed by a competing coherent interaction. For this reason, electrostatic cancellation is treated here as a stress test for gravity-mediated-entanglement protocols, not as a small background detail.

This perspective is useful for experimentalists in a practical way. The results do not claim to provide a full device-level feasibility proposal. Instead, they identify where the useful channel window appears, how it shrinks with finite wave-packet width and dephasing, and where residual electrostatic phases can close the window entirely. The cancellation line gives a simple warning: it is not enough to reduce electrostatic forces only in magnitude; their coherent branch-phase contribution must also be calibrated. The discrimination analysis gives a related message. Near a calibrated mimic point, the gravitational and electrostatic output states may require many independent channel uses to distinguish. Thus, the analysis provides tolerance boundaries and diagnostic tests that can guide the design of future massive-superposition experiments.

The paper is organized as follows. Section~\ref{sec:framework} introduces the finite-width branch-channel model, the correlated dephasing model, the operational witnesses, the benchmark channels, and the electrostatic-contamination model. Section~\ref{sec:Results} presents the Bell--teleportation window, the comparison with local and stochastic reference channels, the electrostatic cancellation band, the metrological and discrimination tests, and the resource-fracture phase diagram. Section~\ref{sec:Conclusion} gives the conclusion, while the Appendix~\ref{app:numerics} collects the numerical implementation details.

\section{Finite-width channel model and operational tests}
\label{sec:framework}

\begin{figure}[t]
    \centering
    \includegraphics[width=1\linewidth]{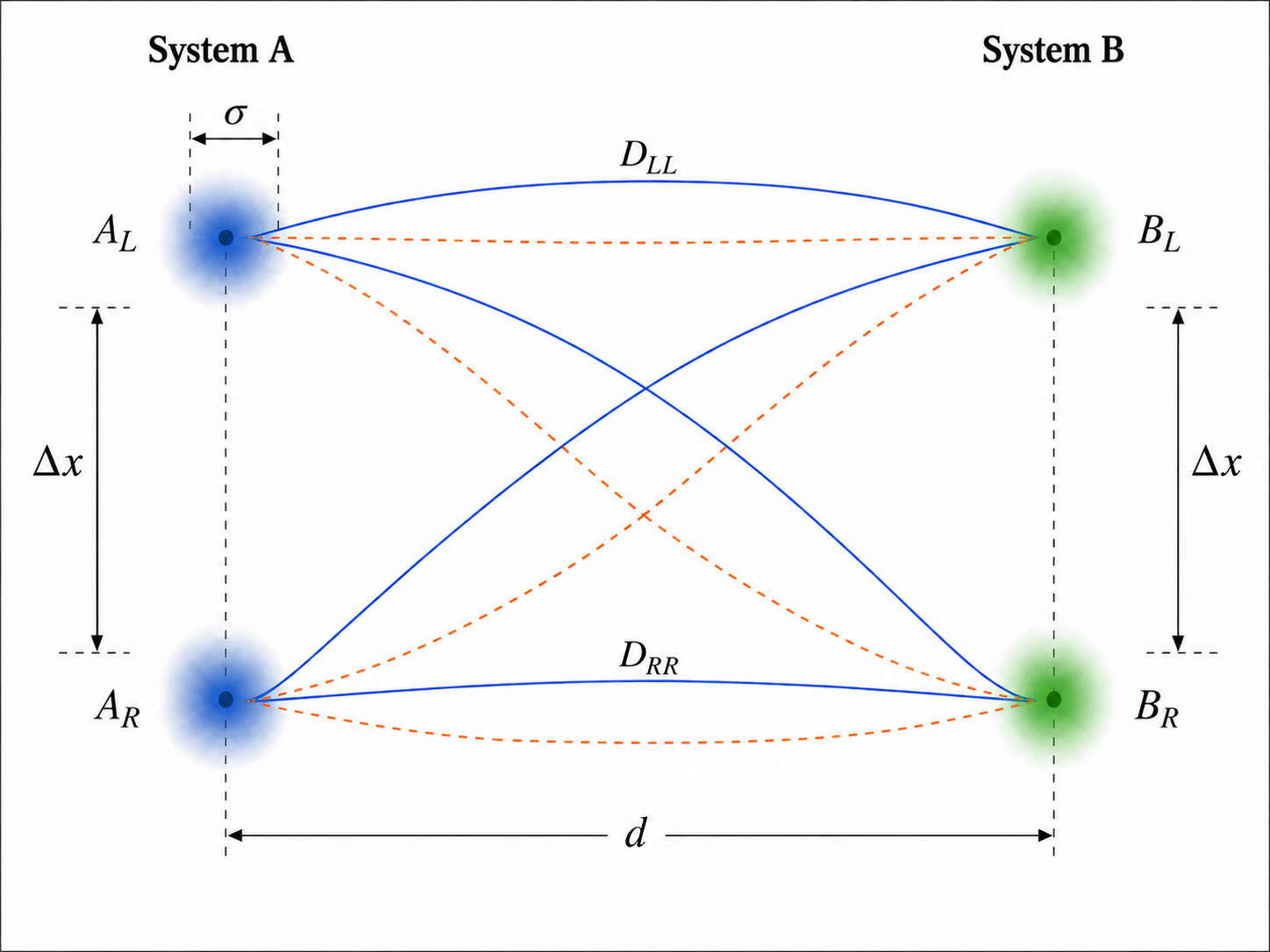}
\caption{
Schematic of the finite-width gravitational cat-state model. Two massive
systems, \(A\) and \(B\), are prepared in spatial superpositions of two
branches labelled \(L\) and \(R\). These symbols are branch labels and do
not represent geometric lengths. The intra-interferometer separation
between the two branches of each system is denoted by \(\Delta x\), while
\(d\) denotes the center-to-center separation between the two
interferometers. The distance between branch \(a\) of system \(A\) and
branch \(b\) of system \(B\) is
\(D_{ab}=|\mathbf{x}_{Aa}-\mathbf{x}_{Bb}|\), with
\(a,b\in\{L,R\}\); the labels \(D_{LL}\) and \(D_{RR}\) shown in the
schematic identify two representative same-label branch separations. The
solid blue and dashed orange curves are schematic interaction links, not
physical propagation paths, and denote gravitational and residual
electrostatic couplings, respectively. The finite wave-packet width
\(\sigma\) modifies the branch-dependent interaction phases and therefore
affects the operational windows for Bell violation, teleportation,
metrology, and interaction discrimination.
}
\label{fig:schematic_model}
\end{figure}

This section defines the model and the diagnostics used in the rest of the paper. The physical picture is simple.
Figure~\ref{fig:schematic_model} gives a schematic view of the model: two massive systems are placed in two-branch spatial superpositions, the finite-width branch wave packets interact gravitationally, and a residual electrostatic background may act on the same branch geometry.
Each system therefore has two possible branches, labelled \(L\) and \(R\). The pair of systems can then be treated as an effective two-qubit system in the branch basis. Gravity assigns different phases to different branch pairs because the branch-to-branch separations vary. If these phases contain a genuinely nonlocal part, they can generate entanglement and other operational advantages.

The model keeps three ingredients. First, the wave packets are not treated as point particles; each branch has a finite Gaussian width. Second, the gravitational interaction produces a branch-dependent phase. Third, environmental noise reduces the coherence between different branch configurations. We intentionally kept the model minimal so that the role of finite width, dephasing, and electrostatic contamination can be clearly seen. Thermal preparation, weak-measurement filtering, and detailed platform-dependent imperfections are not included in the main calculation. The approach follows the spirit of the BMV proposal, but the emphasis here is not only on whether entanglement is generated. We also ask whether the generated state remains useful for Bell tests, teleportation, metrology, and interaction discrimination \cite{Bose2017,MarlettoVedral2017,AnastopoulosHu2015,AnastopoulosHu2020,Rahman2025PRD}.

\subsection{Finite-width gravitational branch channel}
\label{subsec:finite_width_model}

We consider two massive systems, denoted by \(A\) and \(B\). Each system
is prepared in a coherent superposition of two localized branches. We use
\(a,b\in\{L,R\}\) as discrete branch labels. The symbols \(L\) and \(R\)
distinguish the two logical branches and do not denote fixed spatial
directions or geometric distances. The corresponding two-system branch basis is the four-dimensional set
\begin{equation}
    \left\{
    \ket{LL},\ket{LR},\ket{RL},\ket{RR}
    \right\}.
    \label{eq:framework_branch_basis}
\end{equation}
This is the usual effective two-qubit description used in
gravity-mediated-entanglement proposals. In practice, the branch degree
of freedom may be controlled by spin-dependent interferometry, levitated
particles, or related matter-wave platforms
\cite{Bose2017,MarlettoVedral2017,RomeroIsart2011,Pino2018,
Pedernales2020,Neumeier2024,RodaLlordes2024}.

The two branches within each interferometer are separated by
\(\Delta x\), while \(d\) denotes the separation between the centers of
the two interferometers, as shown in Fig.~\ref{fig:schematic_model}. Let
\(\mathbf{x}_{Aa}\) and \(\mathbf{x}_{Bb}\) denote the centers of branch
\(a\) of system \(A\) and branch \(b\) of system \(B\), respectively. The corresponding
branch-pair distance is
\begin{equation}
    D_{ab}
    =
    \left|
    \mathbf{x}_{Aa}-\mathbf{x}_{Bb}
    \right|,
    \qquad a,b\in\{L,R\}.
    \label{eq:framework_branch_distance}
\end{equation}
For the numerical calculations, a convenient coordinate convention is
\begin{align}
    \mathbf{x}_{AL}
    &=
    \left(-\frac{\Delta x}{2},0,0\right),
    &
    \mathbf{x}_{AR}
    &=
    \left(+\frac{\Delta x}{2},0,0\right),
    \nonumber\\
    \mathbf{x}_{BL}
    &=
    \left(-\frac{\Delta x}{2},d,0\right),
    &
    \mathbf{x}_{BR}
    &=
    \left(+\frac{\Delta x}{2},d,0\right).
    \label{eq:framework_branch_coordinates}
\end{align}
Consequently,
\begin{equation}
    D_{LL}=D_{RR}=d,
    \qquad
    D_{LR}=D_{RL}
    =
    \sqrt{d^2+(\Delta x)^2}.
    \label{eq:framework_branch_distances}
\end{equation}
The orientation of the coordinate axes is conventional; only these
branch-pair distances enter the numerical model.

The important difference from the ideal point-particle description is
that each branch is assigned a finite Gaussian width \(\sigma\). We use
\(\sigma\) as the width parameter appearing in the isotropic
Gaussian-smeared kernel and treat \(\ket{L}\) and \(\ket{R}\) as
orthogonal logical branch states throughout the calculation. The broad
upper part of the width scan is therefore a finite-width stress test of
this effective two-level model. To include the spatial extension, we use
the Gaussian-smeared Newtonian kernel
\begin{equation}
    K(D,\sigma)
    =
    \frac{1}{D}
    \operatorname{erf}
    \left(
    \frac{D}{2\sigma}
    \right),
    \qquad D>0,
    \label{eq:framework_kernel}
\end{equation}
where \(D\) is the generic distance argument and
\(\operatorname{erf}(x)\) is the error function. In the physical branch
geometry, this argument is evaluated at one of the four distances
\(D_{ab}\). When \(D\gg\sigma\), the error function approaches unity and
the point-particle kernel \(K(D,\sigma)\rightarrow 1/D\) is recovered.
When \(D\) becomes comparable to \(\sigma\), the short-distance
interaction is softened.

Using the gravitational energy scale \(E_0\) defined in Eq.~\eqref{eq:framework_energy_scale}, the dimensionless
gravitational phase-accumulation rate associated with branch pair \(ab\)
is
\begin{equation}
    \omega_{ab}^{G}
    =
    -d\,K(D_{ab},\sigma_G).
    \label{eq:framework_grav_rate}
\end{equation}
Here \(\sigma_G\) is the same generic width \(\sigma\) in Eq.~\eqref{eq:framework_kernel}, now labeled explicitly as the width of the massive (gravitating) wave packet. The minus sign represents the attractive Newtonian interaction. In the
numerical code, all lengths are measured in units of \(d\) and \(d=1\),
so Eq.~\eqref{eq:framework_grav_rate} is evaluated there as
\(-K(D_{ab},\sigma_G)\). Thus,
\(\omega_{ab}^{G}\) specifies a dimensionless relative
phase-accumulation rate rather than a spatial coordinate or distance.

Some contributions to the four branch phases can be removed by local
rephasings of the two branch qubits. The local-phase-invariant
gravitational rate is
\begin{equation}
    \Omega_G
    =
    \omega_{LL}^{G}
    +
    \omega_{RR}^{G}
    -
    \omega_{LR}^{G}
    -
    \omega_{RL}^{G}.
    \label{eq:framework_nonlocal_rate}
\end{equation}
Thus, \(\Omega_G\) isolates the phase rate associated with the joint
two-body geometry rather than with either interferometer separately. The
corresponding nonlocal phase is
\begin{equation}
    \Phi_G(\tau)
    =
    \Omega_G\tau,
    \label{eq:framework_nonlocal_phase}
\end{equation}
where \(\tau\) is the dimensionless interaction time. If this local-phase-invariant contribution is large enough and coherence is not lost too quickly, the branch state can become Bell-nonlocal or useful for teleportation. The same phase combination is also the one that can be reduced or cancelled by a residual electrostatic interaction. Related nonlocal phase structures appear in gravity-mediated-entanglement analyses and in discussions of whether entanglement can certify nonclassical mediation \cite{Belenchia2018,ChristodoulouRovelli2019,Danielson2022,Carney2022,MartinMartinez2023}.

\subsection{Physical scale and scope of the dimensionless model}
\label{subsec:scope_scaling}

Most of the numerical results are shown in dimensionless variables. This makes the plots general, but it is important to know what the variables mean physically. We define the gravitational energy scale
\begin{equation}
    E_0=\frac{Gm_A m_B}{d},
    \label{eq:framework_energy_scale}
\end{equation}
where \(m_A\) and \(m_B\) are the masses of systems \(A\) and \(B\), and \(t\) denotes physical (dimensional) time, and introduce
\begin{equation}
    \tau=\frac{E_0 t}{\hbar},
    \qquad
    s=\frac{\sigma_G}{d},
    \qquad
    \Gamma_0=\frac{\gamma_0\hbar}{E_0}.
    \label{eq:framework_scaled_variables}
\end{equation}
Here \(\tau\) is the dimensionless time, \(s\) is the wave-packet width measured relative to the interferometer separation, \(\gamma_0\) is the physical (dimensional) dephasing rate, and \(\Gamma_0\) is its dimensionless counterpart.

For equal masses \(m_A=m_B=m\), the physical time corresponding to \(\tau=1\) is
\begin{equation}
    t(\tau=1)
    =
    \frac{\hbar d}{Gm^2}
    \simeq
    1.58~{\rm s}
    \left(\frac{d}{100~\mu{\rm m}}\right)
    \left(\frac{10^{-14}~{\rm kg}}{m}\right)^2 .
    \label{eq:physical_time_scale}
\end{equation}
This scaling shows why the mass is so important. If the mass is reduced by two orders of magnitude, the same dimensionless time can correspond to days instead of seconds. Therefore, the plots below should not be read as direct experimental feasibility claims. They show the channel mechanism and the parameter sensitivity that a real experiment would have to reach.

For reference, Table~\ref{tab:physical_scale} gives representative time and
dephasing scales for equal masses, using \(d=100~\mu{\rm m}\) and
\(\Gamma_0=0.018\).
\begin{table}[t]
\caption{Physical scale associated with the dimensionless time used in the figures for equal masses and \(d=100~\mu{\rm m}\). The last column gives the approximate physical dephasing rate \(\gamma_0=\Gamma_0E_0/\hbar\) for \(\Gamma_0=0.018\).}
\label{tab:physical_scale}
\begin{ruledtabular}
\begin{tabular}{cccc}
\(m\) & \(t(\tau=1)\) & \(t(\tau=18)\) & \(\gamma_0\) \\
\hline
\(10^{-14}\,\mathrm{kg}\) & \(1.58\,\mathrm{s}\) & \(28.4\,\mathrm{s}\) & \(1.1\times10^{-2}\,\mathrm{s}^{-1}\) \\
\(10^{-15}\,\mathrm{kg}\) & \(158\,\mathrm{s}\) & \(47.4\,\mathrm{min}\) & \(1.1\times10^{-4}\,\mathrm{s}^{-1}\) \\
\(10^{-16}\,\mathrm{kg}\) & \(4.39\,\mathrm{h}\) & \(3.29\,\mathrm{d}\) & \(1.1\times10^{-6}\,\mathrm{s}^{-1}\) \\
\(10^{-17}\,\mathrm{kg}\) & \(18.3\,\mathrm{d}\) & \(0.90\,\mathrm{yr}\) & \(1.1\times10^{-8}\,\mathrm{s}^{-1}\)
\end{tabular}
\end{ruledtabular}
\end{table}

The width variable should be interpreted in the same scaled way. For example, if \(d=100~\mu{\rm m}\), then \(0.04\leq s\leq0.55\) corresponds to \(4~\mu{\rm m}\leq\sigma_G\leq55~\mu{\rm m}\). The upper part of this range is intentionally broad. It is included as a stress test of finite-width effects rather than as a claim that such a regime is immediately achievable. A full platform analysis would also need trap geometry, wave-packet expansion, motional heating, residual gas, black-body radiation, charge neutralization, shielding, and readout efficiency. The present work is therefore a finite-width channel study in scaled variables.

\subsection{Spatially correlated dephasing}
\label{subsec:dephasing_model}

The initial branch state is taken to be the product superposition
\begin{equation}
    \rho(0)=\ket{++}\bra{++},
    \qquad
    \ket{+}=\frac{\ket{L}+\ket{R}}{\sqrt{2}}.
    \label{eq:framework_initial_state}
\end{equation}
This choice means that, before the interaction, each mass is coherently delocalized between its two branches, but the two masses are not yet entangled.

As the system evolves, two things happen at the same time. The branch-dependent interaction gives relative phases, while the environment reduces coherence between different branch configurations. We model the latter effect through a spatially correlated dephasing matrix. This is important because massive-superposition experiments are sensitive not only to the strength of local noise, but also to whether the noise seen by the two interferometers is correlated \cite{Kafri2014,TilloyDiosi2016,TilloyDiosi2017,Haddadi2024EPJC}.

Let us write a composite branch index as \(i=(a_i,b_i)\), with \(a_i,b_i\in\{L,R\}\). We use \(\omega_i=\omega_{a_i b_i}^{G}\) for pure-gravity and mean-field calculations, and \(\omega_i=\omega_{a_i b_i}^{\rm tot}\) when electrostatic contamination is included. The evolved density matrix element between the configurations \(i\) and \(j\) is written as
\begin{equation}
    \rho_{ij}(\tau)
    =
    \rho_{ij}(0)
    \exp[-i(\omega_i-\omega_j)\tau]
    \exp[-\Gamma_0 Q_{ij}\tau].
    \label{eq:framework_branch_evolution}
\end{equation}
The first exponential is the coherent phase evolution. The second exponential is the loss of coherence. Thus, Eq.~\eqref{eq:framework_branch_evolution} separates reversible phase accumulation from irreversible dephasing.
The dimensionless factor \(Q_{ij}\) determines how strongly the coherence between branch configurations \(i\) and \(j\) is suppressed. For \(i=(a_i,b_i)\) and \(j=(a_j,b_j)\), we define
\begin{align}
    Q_{ij}
    &=
    \frac{
    |\Delta\mathbf x_A^{ij}|^2
    +
    |\Delta\mathbf x_B^{ij}|^2
    +
    2e^{-d/\ell_c}
    \Delta\mathbf x_A^{ij}\cdot\Delta\mathbf x_B^{ij}
    }
    {(\Delta x)^2},
    \label{eq:framework_Qij}
\end{align}
 with
 \begin{align}
    \Delta\mathbf x_A^{ij}
    &=
    \mathbf x_{A a_i}-\mathbf x_{A a_j},
    \qquad
    \Delta\mathbf x_B^{ij}
    =
    \mathbf x_{B b_i}-\mathbf x_{B b_j}.
    \label{eq:framework_branch_displacements}
\end{align}
Here, \(\Delta x\) is the branch separation, \(d\) is the separation between the two interferometers as before, and \(\ell_c\) is the spatial correlation length of the noise. The construction gives \(Q_{ii}=0\) and \(Q_{ij}=Q_{ji}\).
Because the exponent is obtained from the variance of correlated Gaussian local phases with a positive-semidefinite covariance matrix, the coherence matrix \(\exp[-\Gamma_0Q_{ij}\tau]\) defines the corresponding completely positive dephasing channel. Configurations that differ little in space have smaller \(Q_{ij}\) and retain coherence longer, whereas more widely separated configurations dephase more rapidly.

This form can be understood from Gaussian phase noise. If the noise adds random branch phases \(\delta\phi_i\), then averaging the random phase factor gives
\begin{equation}
    \left\langle e^{-i(\delta\phi_i-\delta\phi_j)}\right\rangle
    =
    \exp\left[-\frac{1}{2}\mathrm{Var}(\delta\phi_i-\delta\phi_j)\right].
\end{equation}
A spatial covariance proportional to \(e^{-d/\ell_c}\) then leads to a distance-dependent decay factor of the form used in Eq.~\eqref{eq:framework_Qij}. In this paper, \(\ell_c\) is not tied to one specific microscopic bath. It represents the effective correlation length of the dominant technical or environmental phase noise. A device-level model would have to derive \(\Gamma_0\) and \(\ell_c\) from concrete sources such as trap noise, thermal radiation, gas scattering, field gradients, or vibrations.

This simplified treatment is useful because it keeps the logic of the paper clean. The same finite-width branch state is used to compute Bell nonlocality, teleportation fidelity, metrological sensitivity, channel discrimination, and resource fracture. Additional preparation and control mechanisms can be added later, but they are not needed to identify the finite-width and electrostatic effects studied here \cite{HallReginatto2018,AnastopoulosHu2018,KentPitalua2021,Galley2022,Galley2023}.

\subsection{Operational communication and correlation witnesses}
\label{subsec:operational_witnesses}

We now define the quantities used to decide whether the generated state is useful. The key point is that entanglement alone is not the strongest operational test. A state may be entangled and still fail to violate a Bell inequality or fail to improve teleportation beyond the classical limit. For this reason, we compare several witnesses.

The first test is Bell nonlocality. From the output state \(\rho(\tau)\), we compute the Pauli-correlation matrix
\begin{equation}
    T_{mn}
    =
    \operatorname{Tr}
    [\rho(\tau)\sigma_m\otimes\sigma_n],
    \qquad
    m,n\in\{x,y,z\}.
    \label{eq:framework_Tmn}
\end{equation}
This matrix contains the two-qubit correlations that would be measured through local Pauli settings. The maximum CHSH value is
\begin{equation}
    \mathcal{B}_{\max}
    =
    2\sqrt{u_1+u_2},
    \label{eq:framework_chsh}
\end{equation}
where \(u_1\) and \(u_2\) are the two largest eigenvalues of \(T^{\rm T}T\). We use \(\mathcal{B}_{\max}\equiv\mathcal{B}_{\rm CHSH}^{\max}\) throughout the figures. The condition
\(\mathcal{B}_{\max}>2\)
indicates Bell-nonlocal correlations for a two-qubit state \cite{Horodecki1995}. This is stronger than simply asking whether the state is entangled.

The second test is teleportation usefulness. The optimal qubit teleportation fidelity is
\begin{equation}
    F_{\rm tel}^{\rm opt}
    =
    \frac{1}{2}
    \left[
    1+
    \frac{s_1+s_2+s_3}{3}
    \right],
    \label{eq:framework_ftel}
\end{equation}
where \(s_1,s_2,s_3\) are the singular values of \(T\), using the standard Horodecki-criterion notation. The classical threshold is
    $F_{\rm tel}^{\rm class}=2/3$.
Thus, \(F_{\rm tel}^{\rm opt}>2/3\) means that the state is useful as a teleportation resource beyond what is classically possible~\cite{Bennett1993,Horodecki1999}.

For comparison with ordinary two-qubit entanglement, we also compute the concurrence
\begin{equation}
    C(\rho)
    =
    \max\{0,\lambda_1-\lambda_2-\lambda_3-\lambda_4\},
    \label{eq:framework_concurrence}
\end{equation}
where \(\lambda_i\) are the ordered square roots of the eigenvalues of
\(\rho(\sigma_y\otimes\sigma_y)\rho^*(\sigma_y\otimes\sigma_y)\) \cite{Wootters1998}. The ordering convention is \(\lambda_1\geq\lambda_2\geq\lambda_3\geq\lambda_4\). Concurrence tells us whether the state is entangled, while Bell violation and teleportation fidelity tell us whether that entanglement is strong enough for stricter tasks.

We also use the one-way coherent information
\begin{equation}
    I_c(A\rangle B)
    =
    S(\rho_B)-S(\rho_{AB}),
    \label{eq:framework_Ic}
\end{equation}
where \(S(\rho)=-\operatorname{Tr}(\rho\log_2\rho)\) is the von Neumann entropy. Note that \(\rho_{AB}=\rho\) is the two-system output state and \(\rho_B=\operatorname{Tr}_A\rho_{AB}\) is the reduced state of subsystem \(B\). Positive coherent information is a stronger indicator that the state has transmission-oriented value \cite{Lloyd1997,Devetak2005}. We do not claim that \(I_c\leq0\) means the state is useless for every task. Instead, we use \(I_c>0\) as a conservative extra requirement when we want the same state to support several operational advantages at once.

\subsection{Model-exclusion benchmarks}
\label{subsec:framework_model_exclusion}
The next question is whether the finite-width gravitational signal could be reproduced by simpler non-gravitational or local models. To address this, we compare the gravitational channel with two reference channels. These benchmarks are not meant to rule out every possible classical or semiclassical theory. They are used to test whether the same operational behavior can be reproduced by two natural alternatives: a local mean-field phase channel and a correlated stochastic local phase channel.

The first benchmark is a local mean-field phase channel. It is obtained by fitting the four gravitational branch rates to the closest additive local form,
\begin{equation}
    \omega_{ab}^{\rm MF}=\omega_0+\alpha_a+\beta_b.
    \label{eq:framework_mf}
\end{equation}
Here, $\omega_0$ is the branch-independent global phase rate,  $\alpha_a$ and $\beta_b$ are the local phase contributions associated
with branch $a$ of system $A$, and so for system $B$. To make the decomposition unique, we use the zero-sum
convention
\begin{equation}
\sum_{a=L,R}\alpha_a=0,
\qquad
\sum_{b=L,R}\beta_b=0,
\end{equation}
then
$\bar{\omega}_{a\cdot}
=
\frac{1}{2}\sum_b\omega_{ab}^{G}$, 
$\bar{\omega}_{\cdot b}
=
\frac{1}{2}\sum_a\omega_{ab}^{G}$, 
and
$\bar{\omega}
=
\frac{1}{4}\sum_{a,b}\omega_{ab}^{G}$,
while the least-squares additive projection is obtained by choosing
$\omega_0 =\bar{\omega}$,
$\alpha_a =\bar{\omega}_{a\cdot}-\bar{\omega}$,
and
$\beta_b =\bar{\omega}_{\cdot b}-\bar{\omega}$.
Consequently, we get
\begin{align}
\omega_{ab}^{\rm MF}
=
\bar{\omega}_{a\cdot}
+
\bar{\omega}_{\cdot b}
-
\bar{\omega}.
\end{align}
Therefore, the term $\omega_0$ contributes only a common global phase, whereas
$\alpha_a$ and $\beta_b$ generate local branch-dependent phases.  This expression contains a global term and one local contribution from each subsystem. It therefore has no genuinely nonlocal branch phase. Indeed,
\begin{equation}
    \omega_{LL}^{\rm MF}+\omega_{RR}^{\rm MF}-\omega_{LR}^{\rm MF}-\omega_{RL}^{\rm MF}=0.
\end{equation}
This comparison is useful because local phase shifts can change interference patterns, but they cannot by themselves generate the same nonlocal branch resource. The mean-field channel is therefore a controlled way to remove the nonlocal component while keeping the same branch geometry~\cite{HallReginatto2018,AnastopoulosHu2018,KentPitalua2021}.

The second benchmark is an optimized Gaussian stochastic separable channel. This model asks whether correlated random local phases, with no nonlocal unitary branch phase, can reproduce the same operational window. The channel is written as
\begin{equation}
    \mathcal{C}_{\rm st}[\rho]
    =
    \mathbb{E}_{\theta_A,\theta_B}
    [U_A(\theta_A)\otimes U_B(\theta_B)\rho
    U_A^\dagger(\theta_A)\otimes U_B^\dagger(\theta_B)],
    \label{eq:framework_stochastic}
\end{equation}
with
    $U_A(\theta_A)=e^{-i\theta_A\sigma_z/2}$
and   
    $U_B(\theta_B)=e^{-i\theta_B\sigma_z/2}$.
Note that we denote these local stochastic phases by \(\theta_A,\theta_B\) rather than \(\xi\), reserving \(\xi\) for the quantum Chernoff exponent \(\xi_{\rm QCB}\) introduced in Sec.~\ref{subsec:framework_metrology_discrimination}. The variables \(\theta_A\) and \(\theta_B\) are zero-mean correlated Gaussian phases with
\begin{equation}
    \operatorname{Var}(\theta_A)
    =
    \operatorname{Var}(\theta_B)
    =
    2\Gamma_0\tau,
    \qquad
    \operatorname{Cov}(\theta_A,\theta_B)
    =
    2\eta\Gamma_0\tau.
    \label{eq:framework_stochastic_covariance}
\end{equation}
Moreover,  $\mathbb{E}_{\theta_A,\theta_B}$ is the ensemble average over the joint probability distribution of the random local phases $\theta_{A, B}$. These phases are selected to be zero-mean jointly Gaussian variables. The parameter $\eta$ is their dimensionless correlation coefficient, defined by
\begin{equation}
\eta
=
\frac{
\operatorname{Cov}(\theta_A,\theta_B)
}{
\sqrt{
\operatorname{Var}(\theta_A)
\operatorname{Var}(\theta_B)
}
}.
\end{equation}

Thus, $\eta=0$ corresponds to uncorrelated local phase fluctuations, $\eta>0$ to positively correlated fluctuations, and $\eta<0$ to anticorrelated fluctuations. The allowed covariance matrix requires $|\eta|\leq1$; in the numerical calculations, we scan the slightly restricted interval $-0.98\leq\eta\leq0.98$.
 Equivalently, the numerical code evaluates
\begin{equation}
    \rho_{ij}^{\rm st}(\tau;\eta)
    =
    \rho_{ij}(0)
    \exp[-\Gamma_0Q_{ij}^{\rm st}(\eta)\tau],
    \label{eq:framework_stochastic_state}
\end{equation}
where, for the logical coordinates \(x_L=-1/2\) and \(x_R=+1/2\), and with
\(\delta x_A^{ij}=x_{a_i}-x_{a_j}\) and \(\delta x_B^{ij}=x_{b_i}-x_{b_j}\)
(the scalar, logical-coordinate analogues of \(\Delta\mathbf x_A^{ij}\) and \(\Delta\mathbf x_B^{ij}\) in Eq.~\eqref{eq:framework_branch_displacements}), we have
\begin{equation}
    Q_{ij}^{\rm st}(\eta)
    =
    (\delta x_A^{ij})^2
    +
    (\delta x_B^{ij})^2
    +
    2\eta\,\delta x_A^{ij}\delta x_B^{ij}.
    \label{eq:framework_Qst}
\end{equation}
This stochastic matrix replaces, rather than supplements, the spatial
dephasing matrix \(Q_{ij}\) used for the gravitational and mean-field
channels. Each realization is local, and the averaged channel is a
convex mixture of local unitaries; therefore it cannot create
entanglement from the separable input.

The Bell and teleportation benchmarks are separate pointwise upper
envelopes,
\begin{equation}
    \mathcal{B}_{\rm st}^{\rm opt}
    =
    \max_\eta\mathcal{B}_{\rm st}(\eta),
    \qquad
    F_{\rm st}^{\rm opt}
    =
    \max_\eta F_{\rm st}(\eta),
    \label{eq:framework_stochastic_envelopes}
\end{equation}
so their maximizing values of \(\eta\) need not coincide. This
comparison excludes only the specified Gaussian random-local-unitary
separable family, not all classical or hybrid models
\cite{Kafri2014,TilloyDiosi2016,TilloyDiosi2017,Galley2022,Galley2023}.

\subsection{Electrostatic contamination as coherent phase cancellation}
\label{subsec:framework_electrostatic}
Residual electrostatic interactions are one of the most important systematic effects in massive-superposition experiments. In this paper, we treat them as branch-dependent finite-width interactions acting on the same geometry as the gravitational channel. The kernel below is a phenomenological smeared Coulomb-like contamination model. It directly represents a residual charge interaction; patch fields, induced dipoles, and shield-modified fields motivate possible profile mismatch but are not claimed to share this microscopic distance law.

The dimensionless electrostatic branch rate is written as
\begin{equation}
    \omega_{ab}^{E}=+d\,K(D_{ab},\sigma_E),
    \qquad
    \kappa_E=\frac{\sigma_E}{\sigma_G}.
    \label{eq:framework_elec_rate}
\end{equation}
The positive sign corresponds to a repulsive interaction in our dimensionless convention. As for the gravitational rate, the numerical code sets \(d=1\), so it evaluates this expression as \(+K(D_{ab},\sigma_E)\). Here, \(\sigma_E\) denotes the effective spatial width assigned to the electrostatic interaction profile. The parameter \(\kappa_E\) allows the electrostatic spatial profile to differ from the mass profile. During a width scan, \(\kappa_E\) is held fixed, so \(\sigma_E=\kappa_E\sigma_G\) varies proportionally with \(\sigma_G\); this proportional scaling is part of the phenomenological model.

If the contaminant were an unscreened residual charge, and if its spatial profile matched the mass profile, the dimensionless electrostatic strength would be approximately
\begin{equation}
    \epsilon
    \simeq
    {\cal S}\,
    \frac{q_A q_B/(4\pi\epsilon_0)}{Gm_A m_B},
    \label{eq:epsilon_physical}
\end{equation}
where \({\cal S}\) represents shielding, geometric suppression, or calibration factors. Also, \(q_A\) and \(q_B\) are the residual electric charges of systems \(A\) and \(B\) and \(\epsilon_0\) is the vacuum permittivity. In what follows, \(e\) is the elementary charge and \(Z_A\) and \(Z_B\) are the corresponding dimensionless charge numbers. For equal masses \(m\) and charges \(q_A=Z_Ae\), \(q_B=Z_Be\), we have
\begin{equation}
    \epsilon
    \simeq
    3.5\times10^{10}\,{\cal S}\,Z_AZ_B
    \left(\frac{10^{-14}~{\rm kg}}{m}\right)^2 .
    \label{eq:epsilon_charge_estimate}
\end{equation}
This estimate explains why electrostatic control is so serious. For \(m\sim10^{-14}~{\rm kg}\), even a single uncompensated elementary charge would be far stronger than gravity unless the effective coupling is suppressed by many orders of magnitude. In the present model, \(\epsilon\) should therefore be read as a residual calibrated electrostatic phase scale after neutralization and shielding, not as a bare charge ratio.
The total branch rate is
\begin{equation}
    \omega_{ab}^{\rm tot}
    =
    \omega_{ab}^{G}+\epsilon \omega_{ab}^{E}.
    \label{eq:framework_total_rate}
\end{equation}
Here \(\epsilon>0\) gives a repulsive contribution in the adopted convention, while \(\epsilon<0\) represents an opposite-sign or effectively attractive contribution. The important point is that the electrostatic interaction does not merely add random noise. It contributes coherently to the same branch-dependent phase structure as gravity. Its invariant nonlocal rate is
\begin{equation}
    \Omega_E
    =
    \omega_{LL}^{E}
    +
    \omega_{RR}^{E}
    -
    \omega_{LR}^{E}
    -
    \omega_{RL}^{E},
    \label{eq:framework_nonlocal_rate_E}
\end{equation}
and the total nonlocal rate is
\begin{equation}
    \Omega_{\rm tot}(s,\epsilon)
    =
    \Omega_G(s)+\epsilon\Omega_E(s),
    \label{eq:framework_total_omega}
\end{equation}
where \(s=\sigma_G/d\). Therefore, if the electrostatic contribution has the opposite sign in the invariant phase combination, it can cancel the gravitational signal.
 The cancellation point is
\begin{equation}
    \epsilon_{\rm canc}(s)
    =
    -\frac{\Omega_G(s)}{\Omega_E(s)}.
    \label{eq:framework_eps_cancel}
\end{equation}
At this value, the nonlocal branch phase rate vanishes in the model. This does not mean that gravity has disappeared from the Hamiltonian. It means that the specific nonlocal phase combination used to generate the operational signal has been cancelled by a competing coherent interaction. This is why electrostatic contamination is more dangerous than ordinary dephasing: it can erase the useful phase even when coherence has not completely vanished \cite{Carney2019,Pedernales2020,Neumeier2024,RodaLlordes2024,Vijayan2024}.

To test whether the same state is simultaneously useful for several tasks, we define the strict operational margin
\begin{align}
    \mathcal{O}(\tau;s,\epsilon)
    =
    \min\big\{
    &\mathcal{B}_{\max}(\tau;s,\epsilon)-2,\nonumber\\
    &F_{\rm tel}^{\rm opt}(\tau;s,\epsilon)-\frac{2}{3},
    I_c(\tau;s,\epsilon)
    \big\}.
    \label{eq:framework_operational_margin}
\end{align}
This expression takes the weakest of three requirements: Bell violation, teleportation advantage, and positive coherent information. The optimized margin is
\begin{equation}
    \mathcal{O}_{\max}(s,\epsilon)
    =
    \max_{\tau\in[0.60, 18]}
    \mathcal{O}(\tau;s,\epsilon).
    \label{eq:framework_Omax}
\end{equation}
Note that the interval $0.60\leq\tau\leq18$ is chosen as a common finite-time window for all numerical choice optimizations. The lower bound avoids the trivial initial state, while $\tau=18$ sets a practical upper limit for comparing the different operational quantities consistently. 
 
Thus, \(\mathcal{O}_{\max}>0\) means that all three requirements are satisfied at the same finite interaction time. This is deliberately conservative. If different tasks were allowed to use different optimized times, the useful region could be larger. Here we use the stricter same-state, same-time criterion so that the word ``operational'' has a clear and cautious meaning.

\subsection{Metrological and discrimination quantities}
\label{subsec:framework_metrology_discrimination}
The same finite-width channel can also be viewed as a probe of the gravitational interaction strength. To quantify this, we introduce a dimensionless parameter \(g\) that rescales the gravitational branch rates,
\begin{equation}
    \omega_{ab}^{G}\rightarrow g \omega_{ab}^{G}.
\end{equation}
When differentiating with respect to \(g\), the reference energy \(E_0\) and the numerical variables \(\tau\), \(\Gamma_0\), and \(\epsilon\) are held fixed. If the output state changes strongly when \(g\) changes slightly, then the state carries useful information about the gravitational amplitude.

The quantum Fisher information for the parameter-dependent state \(\rho_g\) is
\begin{equation}
    F_Q^{(g)}
    =
    2\sum_{\mu,\nu}
    \frac{|\langle\mu|\partial_g\rho_g|\nu\rangle|^2}
    {\lambda_\mu+\lambda_\nu},
    \label{eq:framework_qfi}
\end{equation}
where \(\rho_g=\sum_\mu\lambda_\mu\ket{\mu}\bra{\mu}\), and terms with \(\lambda_\mu+\lambda_\nu=0\) are omitted. The corresponding Cram\'{e}r--Rao bound is
\begin{equation}
    \delta g\geq\frac{1}{\sqrt{N F_Q^{(g)}}},
\end{equation}
where \(N\) is the number of repetitions~\cite{Giovannetti2011,Degen2017}. The quantity \(\delta g\) denotes the estimation uncertainty, i.e., the standard deviation of an unbiased estimator of the gravitational-amplitude parameter \(g\). In addition to this ideal quantity, we also compute Fisher information from measurable Pauli correlations. This is useful because full collective measurements may be difficult in near-term experiments, whereas Pauli-correlation measurements are closer to the diagnostics already used for Bell and teleportation tests.

We also ask whether a gravitational output state can be distinguished from a calibrated electrostatic mimic; here \(\rho_G\) and \(\rho_E\) denote, respectively, the output state of the purely gravitational channel and of the calibrated electrostatic-mimic construction detailed in Sec.~\ref{subsec:geometry_calibrated_discrimination}.
\\
The single-copy trace distance is:
\begin{equation}
    D_{\rm tr}(\rho_G,\rho_E)
    =
    \frac{1}{2}\|\rho_G-\rho_E\|_1,
    \label{eq:framework_trace_distance}
\end{equation}
where \(\|A\|_1=\operatorname{Tr}\sqrt{A^\dagger A}\) denotes the trace norm.
which gives the equal-prior success probability
\begin{equation}
    P_{\rm succ}^{(1)}=\frac{1+D_{\rm tr}}{2}.
\end{equation}
A larger trace distance means that the two candidate output states are easier to distinguish in one shot.

For many-copy discrimination, we use the quantum Chernoff exponent (Quantum Chernoff Bound, QCB)
\begin{equation}
    \xi_{\rm QCB}
    =
    -\log\left[
    \min_{0\leq q\leq1}
    \operatorname{Tr}(\rho_G^q\rho_E^{1-q})
    \right],
    \label{eq:framework_qcb}
\end{equation}
where the minimization is over the real Chernoff parameter \(q\in[0,1]\). The asymptotic error probability for \(n\) copies (independent uses of the channel) scales as
\begin{equation}
    P_{\rm err}^{(n)}\sim e^{-n\xi_{\rm QCB}}.
\end{equation}
Thus, a small Chernoff exponent means that many copies are needed to identify the interaction mechanism reliably \cite{Audenaert2007}. In the calibrated-discrimination test used later, the electrostatic mimic is matched at one reference geometry and then tested at other wave-packet widths. This prevents the mimic from being re-tuned at every point and makes the comparison more meaningful.

\subsection{Measurement interpretation and statistical scope}
\label{subsec:measurement_scope}

The Pauli correlations used above should be understood as ideal state-diagnostic observables. In an actual massive-superposition experiment, the branch qubit would have to be mapped to an accessible readout degree of freedom. This could be an internal spin, a path population, or an interferometric phase measurement, followed by local rotations that realize the effective \(\sigma_x\), \(\sigma_y\), and \(\sigma_z\) settings.

The present model does not include detector inefficiency, imperfect rotations, leakage out of the branch subspace, or loophole-free Bell-test constraints. These effects would reduce the observed margins and must be included in a device-level proposal. Therefore, the thresholds used in the figures should be read as ideal operational thresholds, not as direct experimental confidence levels.

A simple statistical estimate can nevertheless be stated. For a Pauli-product measurement \(P=\sigma_m\otimes\sigma_n\) with outcomes \(\pm1\), the estimator of \(\langle P\rangle\) has standard error
\begin{equation}
    \Delta P
    =
    \sqrt{\frac{1-\langle P\rangle^2}{N_P}},
\end{equation}
where \(N_P\) is the number of repetitions for that setting. A fixed CHSH expression is a linear combination of four such correlators, so its uncertainty follows by ordinary error propagation. The optimized CHSH value used in the figures is computed from the reconstructed matrix \(T\). In an experiment, the relevant measurement settings should either be fixed in advance from a calibrated model or the statistical penalty from scanning over \(\tau\) and measurement settings should be included. The same caution applies to teleportation fidelity reconstructed from the singular values of \(T\). A positive experimental claim would require the predicted margin above threshold to exceed both statistical uncertainty and systematic readout errors.

\subsection{Resource-fracture classification}
\label{subsec:framework_fracture}

Finally, we classify the channel into three regimes. This classification is meant to make the main message of the paper visually and conceptually clear: entanglement can survive even after stronger operational advantages are lost.

First, we define the maximum concurrence
\begin{equation}
    C_{\max}(s,\epsilon)
    =
    \max_{\tau\in[0.60,18]} C(\tau;s,\epsilon).
    \label{eq:framework_Cmax}
\end{equation}
The fully operational region is
\begin{equation}
    \mathsf{U}:\quad \mathcal{O}_{\max}>0.
    \label{eq:framework_U}
\end{equation}
In this region, the channel satisfies the strict same-time operational condition.

The numerical classification uses the concurrence tolerance
\begin{equation}
    C_{\min}=10^{-4}.
    \label{eq:framework_Cmin}
\end{equation}
The fractured region is defined by
\begin{equation}
    \mathsf{F}:\quad C_{\max}>C_{\min},
    \qquad
    \mathcal{O}_{\max}\leq0.
    \label{eq:framework_F}
\end{equation}
Here, the state remains entangled above the numerical tolerance but no longer satisfies the stricter joint operational test.

The depleted region is
\begin{equation}
    \mathsf{D}:\quad C_{\max}\leq C_{\min},
    \qquad
    \mathcal{O}_{\max}\leq0.
    \label{eq:framework_D}
\end{equation}
In this region, neither concurrence above the adopted tolerance nor the stricter operational margin survives.

This classification uses a standard fact from quantum information theory: entanglement is not equivalent to every quantum advantage. Many entangled states do not violate a Bell inequality and do not satisfy stronger transmission criteria. The contribution here is to apply this separation to finite-width gravitational cat-state channels under dephasing and electrostatic cancellation. In particular, \(C_{\max}>C_{\min}\) does not imply simultaneous Bell violation, teleportation advantage, and positive coherent information. The resource-fracture analysis therefore separates ordinary correlation survival from useful gravitational-channel behavior \cite{Wootters1998,Horodecki1995,Horodecki1999,Devetak2005,Rahman2025PRD}.

\section{Results and Discussion}
\label{sec:Results}
The results are organized around a practical question: after finite wave-packet width and dephasing are included, when does the gravitational cat state still behave as a useful channel that can be distinguished from the benchmark models considered here? We answer this in stages. First, we locate the time and width intervals where Bell violation and teleportation advantage occur together. We then compare this window with two reference channels, one local and one stochastic-separable, to check whether the same operational behavior appears without the nonlocal branch phase. After that, we add residual electrostatic coupling and follow how it changes communication usefulness, metrological sensitivity, dephasing tolerance, discrimination cost, and the separation between entanglement survival and stronger operational usefulness.

\subsection{Joint Bell--teleportation window of the finite-width gravcat channel}
\label{subsec:joint_window}

We begin with the simplest operational question. After the finite-width gravitational phase and dephasing are included, is there still a time interval in which the same two-branch state is useful for more than one task? To answer this, we require two conditions at the same interaction time: violation of the Bell--CHSH bound and teleportation fidelity above the classical threshold. This is a stricter requirement than observing nonzero concurrence, and it is based on the measurable two-qubit Pauli-correlation matrix.

The finite-width branch interaction is generated by the Gaussian-smeared kernel in Eq.~\eqref{eq:framework_kernel}. In dimensional units, the corresponding gravitational potential is \(V_{ab}^{G}=-Gm_A m_B K(D_{ab},\sigma)\), so the point-particle result \(V_{ab}^{G}\to-Gm_A m_B/D_{ab}\) is recovered when \(\sigma/D_{ab}\to0\). The branch phase is \(\phi_{ab}=V_{ab}^{G}t/\hbar\), while its local-phase-invariant component is the same nonlocal phase \(\Phi_G=\Omega_G\tau\) defined in Eqs.~\eqref{eq:framework_nonlocal_rate} and \eqref{eq:framework_nonlocal_phase}. This phase, rather than a removable single-particle phase, controls the generation of the two-body operational resource.

For the operational test, we use the coherent branch input in Eq.~\eqref{eq:framework_initial_state} and the noisy evolution in Eq.~\eqref{eq:framework_branch_evolution}. The previously undefined branch-distance factor is now fully specified by Eqs.~\eqref{eq:framework_Qij} and \eqref{eq:framework_branch_displacements}; no second definition of \(Q_{ij}\) is required in the Results section. The Bell and teleportation quantities are evaluated from the common correlation matrix and formulas in Eqs.~\eqref{eq:framework_Tmn}--\eqref{eq:framework_ftel}. Thus, the thresholds used below are \(\mathcal{B}_{\max}>2\) and \(F_{\rm tel}^{\rm opt}>2/3\).

\begin{figure*}[t]
    \centering
    \includegraphics[width=\linewidth]{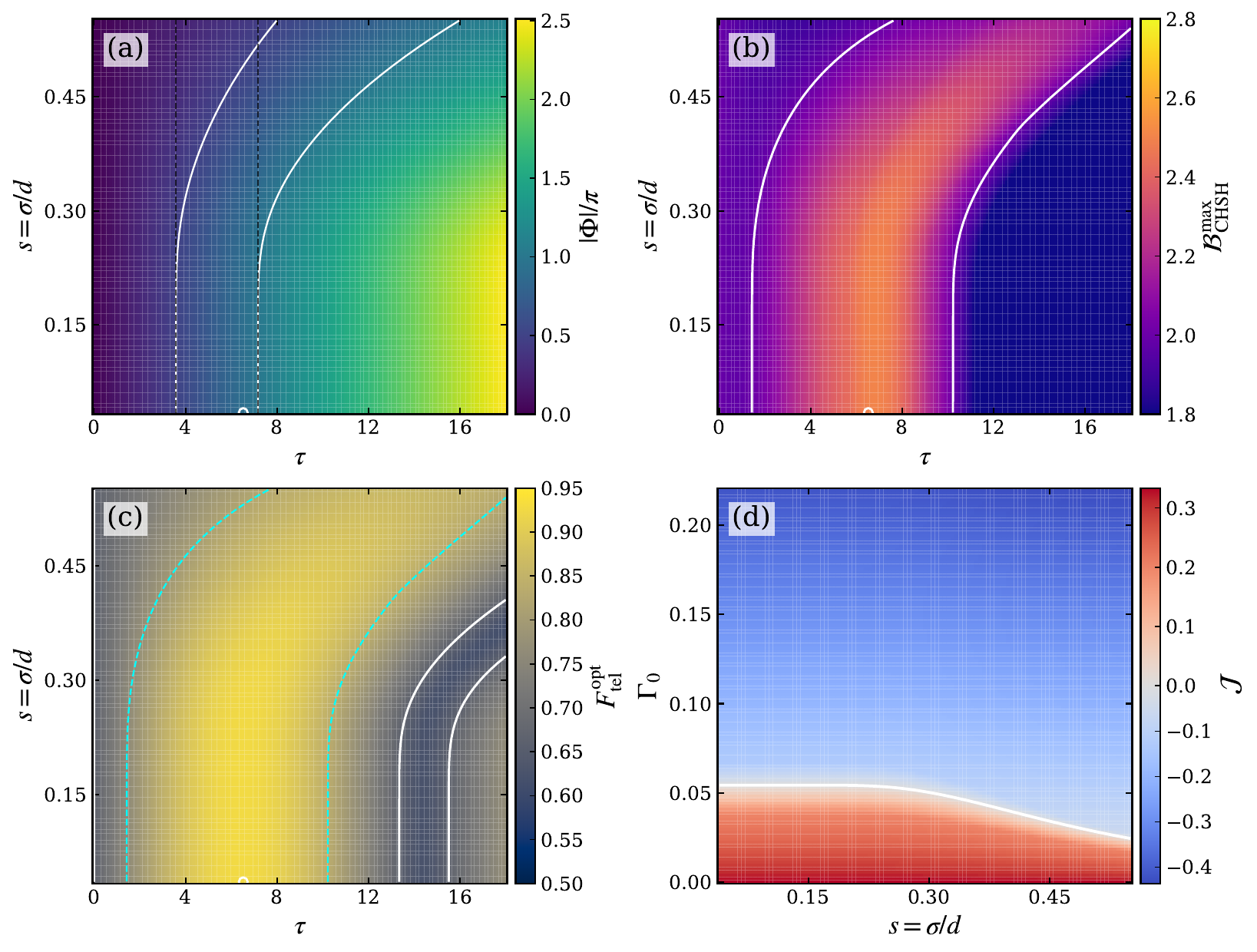}
    \caption{
Finite-width gravitational cat channel and its experimentally testable joint operational window.
The dimensionless variables are \(\tau=tE_0/\hbar\) and \(s=\sigma_G/d\), where \(E_0\) is the gravitational energy scale, \(\sigma_G\) is the Gaussian wave-packet width, and \(d\) is the interferometer separation. The calculations use \(d=1\), \(\Delta x/d=0.80\), \(\hbar=1\), and \(Gm_A m_B/d=1\). In panels (a)--(c), \(\Gamma_0=0.018\) and \(\ell_c/d=1.25\).
(a) Normalized finite-width nonlocal phase \(|\Phi_G|/\pi\). White contours mark \(|\Phi_G|/\pi=0.5\) and \(1\), while black dashed vertical lines mark the point-particle benchmark times.
(b) Maximum Bell--CHSH value \(\mathcal{B}_{\rm CHSH}^{\max}\); the white contour marks \(\mathcal{B}_{\rm CHSH}^{\max}=2\).
(c) Optimal teleportation fidelity \(F_{\rm tel}^{\rm opt}\); the white contour marks \(F_{\rm tel}^{\rm class}=2/3\), while the dashed cyan contour marks \(\mathcal{B}_{\rm CHSH}^{\max}=2\).
(d) Finite-time joint witness \(\mathcal{J}\), evaluated over \(0.60\leq\tau\leq18\), versus \(s\) and \(\Gamma_0\) for fixed \(\ell_c/d=1.25\). The white curve denotes \(\mathcal{J}=0\). The numerical grids are listed in Table~\ref{tab:numerical_grids}.
}
    \label{fig:fig1}
\end{figure*}

The scaled variables are those in Eq.~\eqref{eq:framework_scaled_variables}. Throughout Fig.~\ref{fig:fig1}, we use \(d=1\), \(\Delta x/d=0.80\), \(\hbar=1\), and \(Gm_A m_B/d=1\). For panels~\ref{fig:fig1}(a)--(c), the dephasing parameters are fixed at \(\Gamma_0=0.018\) and \(\ell_c/d=1.25\). The scanned domain is $0\leq \tau\leq18$
with $0.035\leq s\leq0.55$.

To identify the region where both operational advantages occur together, we define the finite-time joint witness
\begin{align}
    \mathcal{J}(s,\Gamma_0)
    =
    \max_{\tau\in[0.60,18]}
    \min
    \bigg\{
    &\mathcal{B}_{\rm CHSH}^{\max}(\tau,s,\Gamma_0)-2,\nonumber\\
    &F_{\rm tel}^{\rm opt}(\tau,s,\Gamma_0)-\frac{2}{3}
    \bigg\}.
    \label{eq:J_definition}
\end{align}
The lower bound \(\tau=0.60\) excludes the trivial initial-time contribution and focuses the optimization on finite interaction times. A positive value, \(\mathcal{J}>0\), means that there exists a finite interaction time at which the same gravitationally generated state both violates the CHSH bound and beats the classical teleportation limit. A negative value, \(\mathcal{J}<0\), means that at least one of the two operational requirements is not satisfied.

Throughout the paper, a subscript or superscript asterisk (e.g.\ \(s_\ast\), \(\tau_\ast\), \(\epsilon_\ast\)) denotes the sampled grid value at which the quantity under discussion in that paragraph is maximized (or, where stated, the value closest to a target); we restate explicitly, at each occurrence, which quantity is being optimized. With this convention, the best sampled operating point in panels~\ref{fig:fig1}(a)-(c), i.e.\ the values of \(s\) and \(\tau\) at which \(\mathcal{J}(s,\Gamma_0)\) of Eq.~\eqref{eq:J_definition} attains its maximum on the scanned grid, together with the corresponding \(\mathcal{B}_{\max}\) and \(F_{\rm tel}^{\rm opt}\) at that point, is
\begin{align}
    &s_\ast=0.0350,
    \quad
    \tau_\ast=6.5235,
    \quad
    \mathcal{J}=0.2596,
    \nonumber\\
    &\mathcal{B}_{\max}=2.4956,
    \quad
    F_{\rm tel}^{\rm opt}=0.9263.
    \label{eq:fig1_best_point}
\end{align}

Figure~\ref{fig:fig1} summarizes the finite-width operational behavior. Panel (a) shows the normalized nonlocal phase \(|\Phi|/\pi\), where \(\Phi\equiv\Phi_G\). The white contours mark \(|\Phi|/\pi=0.5\) and \(1\), while the black dashed vertical lines indicate the corresponding point-particle benchmark times, i.e.\ the times at which the point-particle (\(\sigma_G\to0\)) limit of \(|\Phi_G|/\pi\) itself reaches \(0.5\) and \(1\). The shift away from the dashed benchmark demonstrates that finite wave-packet width modifies the rate of nonlocal gravitational phase accumulation.

Panel (b) shows \(\mathcal{B}_{\rm CHSH}^{\max}\). The white contour marks the threshold \(\mathcal{B}_{\rm CHSH}^{\max}=2\). This panel identifies the finite-width regions in which the gravitational cat channel produces Bell-nonlocal output states. Panel (c) shows \(F_{\rm tel}^{\rm opt}\), with the white contour marking the classical threshold \(F_{\rm tel}^{\rm class}=2/3\). The dashed cyan contour marks \(\mathcal{B}_{\rm CHSH}^{\max}=2\). The overlap between the two threshold regions gives the parameter window in which the generated state is both Bell-nonlocal and teleportation-useful.

Panel (d) gives the robustness test against dephasing. It plots \(\mathcal{J}\) versus \(s=\sigma_G/d\) and \(\Gamma_0\), with fixed \(\ell_c/d=1.25\). The white curve denotes the critical boundary \(\mathcal{J}=0\). Below this boundary, \(\mathcal{J}>0\), and the finite-width gravitational channel supports simultaneous Bell violation and teleportation advantage. Above the boundary, \(\mathcal{J}<0\), and dephasing removes the joint operational usefulness. The downward trend of the boundary indicates that broader wave packets tolerate less dephasing before losing simultaneous Bell and teleportation advantage.

\subsection{Comparison with local and stochastic reference channels}
\label{subsec:model_exclusion}

We next ask whether the operational behavior found above is tied to the nonlocal branch phase or whether it also appears in simpler reference models. This comparison is necessary because local phase shifts and correlated classical phase noise can change branch coherences without providing a genuinely nonlocal channel resource. We therefore compare three channels constructed from the same input and branch geometry. The gravitational and mean-field channels use the spatial dephasing matrix \(Q_{ij}\), whereas the stochastic benchmark uses its own correlated local-phase matrix \(Q_{ij}^{\rm st}(\eta)\) defined in Eq.~\eqref{eq:framework_Qst}.

The first channel is the finite-width quantum gravitational phase channel, denoted by \(\mathcal{G}_Q\). It uses the finite-width gravitational rates in Eq.~\eqref{eq:framework_grav_rate}; its nonlocal part is determined by \(\Omega_G\) in Eq.~\eqref{eq:framework_nonlocal_rate}. This makes explicit that the comparison is performed at the level of the same branch geometry and finite-width interaction used in Fig.~\ref{fig:fig1}.

The second channel is the local mean-field reference \(\mathcal{G}_{\rm MF}\). Its additive form is defined once in Eq.~\eqref{eq:framework_mf}. For numerical implementation, write the full branch-rate matrix as

\begin{equation}
    W_G=
    \begin{pmatrix}
    \omega_{LL}^{G} & \omega_{LR}^{G}\\
    \omega_{RL}^{G} & \omega_{RR}^{G}
    \end{pmatrix}.
    \label{eq:fig2_rate_matrix}
\end{equation}
The least-squares additive projection is
\begin{equation}
    (W_{\rm MF})_{ab}
    =
    \bar\omega_{a\cdot}
    +
    \bar\omega_{\cdot b}
    -
    \bar\omega,
    \label{eq:local_projection_matrix}
\end{equation}

where \(\bar\omega_{a\cdot}\), \(\bar\omega_{\cdot b}\), and \(\bar\omega\) denote the row, column, and total means. This projection has zero invariant nonlocal rate, as stated after Eq.~\eqref{eq:framework_mf}, while retaining the closest additive local phase pattern associated with the same four branch rates.

The third reference is the optimized classical stochastic separable channel \(\mathcal{C}_{\rm st}^{\rm opt}\), defined in Eqs.~\eqref{eq:framework_stochastic}--\eqref{eq:framework_stochastic_envelopes}. The code scans \(-0.98\leq\eta\leq0.98\) and independently constructs the pointwise Bell and teleportation upper envelopes. Each realization is a product of local unitaries, so the averaged map remains separable and cannot create entanglement from the input in Eq.~\eqref{eq:framework_initial_state}. This optimization is favorable within the specified Gaussian random-local-unitary class but does not represent an optimization over all classical or hybrid models.

All three channels use the same initial state and operational diagnostics. The gravitational and mean-field states are propagated with Eq.~\eqref{eq:framework_branch_evolution} using \(Q_{ij}\), while the stochastic state is evaluated with Eq.~\eqref{eq:framework_stochastic_state} using \(Q_{ij}^{\rm st}(\eta)\). Their Bell and teleportation values are then evaluated with Eqs.~\eqref{eq:framework_Tmn}--\eqref{eq:framework_ftel}.

To quantify the simultaneous separation between the quantum gravitational channel and the two reference models, we define the model-exclusion margin
\begin{align}
    \mathcal{X}(\tau)
    =
    \min
    \Bigg\{&
    \mathcal{B}_{Q}(\tau)-2,\,
    F_{Q}^{\rm tel}(\tau)-\frac{2}{3},
    \nonumber\\
    &
    \mathcal{B}_{Q}(\tau)-\mathcal{B}_{\rm MF}(\tau),\,F_{Q}^{\rm tel}(\tau)-F_{\rm MF}(\tau),\nonumber\\
    &\mathcal{B}_{Q}(\tau)-\mathcal{B}_{\rm st}^{\rm opt}(\tau),\,
    F_{Q}^{\rm tel}(\tau)-F_{\rm st}^{\rm opt}(\tau)
    \Bigg\}.
    \label{eq:X_time_definition}
\end{align}
Here \(\mathcal{B}_{Q}\) and \(F_Q^{\rm tel}\) are the Bell and teleportation values of the quantum gravitational channel (the superscript ``tel'' distinguishes this teleportation fidelity from the quantum Fisher information \(F_Q^{(g)}\) of Eq.~\eqref{eq:framework_qfi}), \(\mathcal{B}_{\rm MF}\) and \(F_{\rm MF}\) are the corresponding values for the local mean-field channel, and \(\mathcal{B}_{\rm st}^{\rm opt}\) and \(F_{\rm st}^{\rm opt}\) are the optimized classical stochastic separable benchmarks. A positive value of \(\mathcal{X}(\tau)\) means that, at the same interaction time, the quantum gravitational channel is Bell-nonlocal, teleportation-useful, and above both reference channels. We then optimize this quantity over finite interaction times,
\begin{equation}
    \mathcal{X}(s,\Gamma_0)
    =
    \max_{\tau\in[0.60,18]}
    \mathcal{X}(\tau,s,\Gamma_0).
    \label{eq:X_map_definition}
\end{equation}
The lower bound \(\tau=0.60\) excludes the trivial initial point and focuses on finite-time channel evolution.

\begin{figure*}[t]
    \centering
    \includegraphics[width=\linewidth]{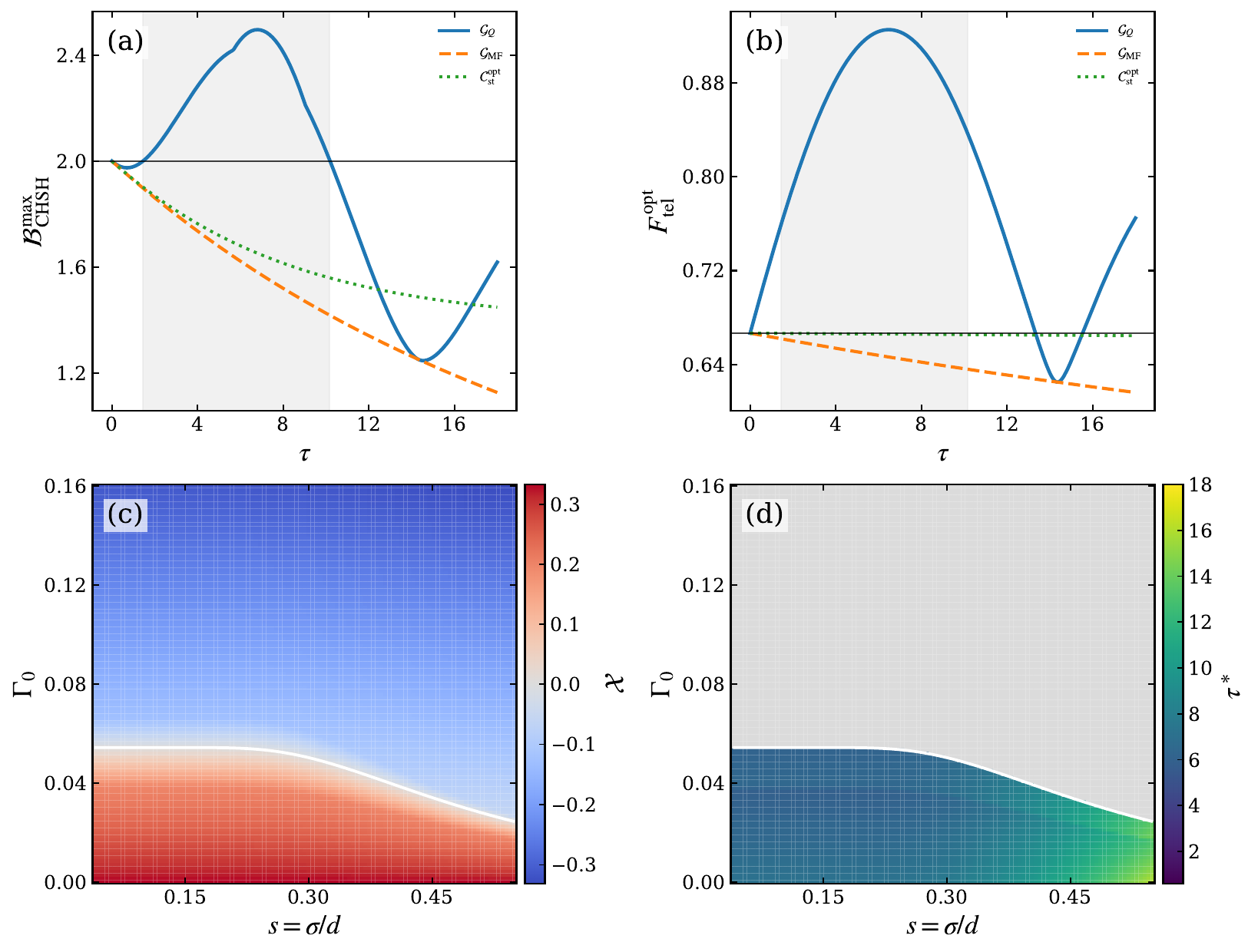}
    \caption{
Operational exclusion of the local and stochastic benchmark channels.
The compared channels are the finite-width quantum gravitational channel \(\mathcal{G}_Q\), the local mean-field projection \(\mathcal{G}_{\rm MF}\), and the Gaussian stochastic separable benchmark \(\mathcal{C}_{\rm st}^{\rm opt}\). The gravitational and mean-field channels use \(Q_{ij}\), whereas the stochastic benchmark uses \(Q_{ij}^{\rm st}(\eta)\). The calculations use \(d=1\), \(\Delta x/d=0.80\), \(\hbar=1\), \(Gm_A m_B/d=1\), and \(\ell_c/d=1.25\).
(a) Maximum Bell--CHSH value versus \(\tau\) at the representative sampled values \(s_\ast=0.0400\) and \(\Gamma_0=0.0182\). The black line marks \(\mathcal{B}_{\rm CHSH}^{\max}=2\).
(b) Optimal teleportation fidelity for the same parameters. The black line marks \(F_{\rm tel}^{\rm class}=2/3\). In panels (a) and (b), the shaded interval denotes \(\mathcal{X}(\tau)>0\), and the full displayed time range is \(0\leq\tau\leq18\).
(c) Optimized model-exclusion margin \(\mathcal{X}(s,\Gamma_0)\), evaluated over \(0.60\leq\tau\leq18\). The white curve denotes \(\mathcal{X}=0\).
(d) Earliest sampled finite interaction time \(\tau_\ast\) maximizing \(\mathcal{X}\) inside the positive-exclusion region. The gray region corresponds to \(\mathcal{X}\leq0\). The stochastic Bell and teleportation curves are separate pointwise upper envelopes over \(\eta\in[-0.98,0.98]\).
}
    \label{fig:fig2}
\end{figure*}

The numerical results in Fig.~\ref{fig:fig2} use the same dimensionless units as Fig.~\ref{fig:fig1}. The map scans
\begin{equation}
    0.04\leq s=\frac{\sigma_G}{d}\leq0.55,
    \qquad
    0\leq \Gamma_0\leq0.16,
\end{equation}
and optimizes over \(0.60\leq\tau\leq18\). Panels (a) and (b) show the full interval \(0\leq\tau\leq18\). The representative sampled dephasing value is \(\Gamma_0=0.0182\) (the nearest point on the 120-point \(\Gamma_0\) grid of Table~\ref{tab:numerical_grids} to the nominal value \(\Gamma_0=0.018\) used elsewhere), and the width that maximizes \(\mathcal{X}\) on that slice is
\begin{equation}
    s_\ast=0.0400,
    \qquad
    \max_s\mathcal{X}(s,~0.0182)=0.2588.
    \label{eq:fig2_representative_point}
\end{equation}

Figure~\ref{fig:fig2} shows the resulting model-exclusion test. Panel (a) compares \(\mathcal{B}_{\rm CHSH}^{\max}\) for \(\mathcal{G}_Q\), \(\mathcal{G}_{\rm MF}\), and \(\mathcal{C}_{\rm st}^{\rm opt}\). Only the finite-width quantum gravitational channel crosses the CHSH threshold, while the local mean-field and optimized classical stochastic separable channels remain below it. Panel (b) shows the corresponding teleportation fidelity. Again, \(\mathcal{G}_Q\) exceeds the classical threshold, whereas the reference channels do not. The shaded interval in panels (a) and (b) marks the finite-time region where \(\mathcal{X}(\tau)>0\), meaning that the quantum gravitational channel is simultaneously useful and separated from both references.
Panel (c) gives the optimized model-exclusion map \(\mathcal{X}(s,\Gamma_0)\). The white contour denotes \(\mathcal{X}=0\). The region \(\mathcal{X}>0\) identifies parameters for which the finite-width quantum gravitational channel violates CHSH, beats the teleportation threshold, and outperforms both the local mean-field channel and the optimized classical stochastic separable channel. The region \(\mathcal{X}<0\) indicates that at least one of these requirements fails. Panel (d) shows the finite interaction time \(\tau_\ast\) at which \(\mathcal{X}\) is maximized inside the positive-exclusion region; when the sampled maximum is attained on more than one grid point (a degenerate maximum), we report the earliest such \(\tau_\ast\), following the convention stated in Appendix~\ref{app:numerics}. The gray region corresponds to \(\mathcal{X}\leq0\), where no finite-time operational exclusion is obtained. This panel shows that the exclusion window is not only a threshold effect, but also has a definite optimal timing structure.

The main observation from Fig.~\ref{fig:fig2} is that the finite-width quantum gravitational cat channel produces an operational signal that is not reproduced by the corresponding local mean-field phase channel or by the optimized Gaussian random-local-unitary separable benchmark considered here. Thus, the useful window identified in Fig.~\ref{fig:fig1} is not only a generic consequence of local dephasing or stochastic phase noise; within this comparison, it is tied to the nonlocal branch phase of the finite-width gravitational interaction.

\subsection{Electrostatic contamination and cancellation}
\label{subsec:electrostatic_tolerance}

We now turn to the most delicate competing interaction: a residual electrostatic force. In an experimentally motivated massive-superposition setting, such a force may coexist with gravity and contribute its own branch-dependent phase. Because this phase enters the same invariant branch combination as the gravitational phase, it can either reinforce the desired signal or partially cancel it. The purpose of this subsection is to track this effect in terms of Bell and teleportation usefulness rather than only through the phase itself.

The gravitational and electrostatic branch rates are those defined in Eqs.~\eqref{eq:framework_grav_rate} and \eqref{eq:framework_elec_rate}, and their coherent sum is the total rate in Eq.~\eqref{eq:framework_total_rate}. Here \(\epsilon=0\) gives the purely gravitational channel, \(\epsilon>0\) denotes a repulsive electrostatic contribution, and \(\epsilon<0\) denotes an opposite-sign or effectively attractive contribution in the adopted convention. The width ratio \(\kappa_E=\sigma_E/\sigma_G\) controls whether the electrostatic profile is spatially matched to, broader than, or narrower than the gravitational mass profile.

The relevant local-phase-invariant quantity is the total nonlocal rate in Eq.~\eqref{eq:framework_total_omega}. Its zero occurs at the cancellation amplitude in Eq.~\eqref{eq:framework_eps_cancel}. At this point the nonlocal phase rate vanishes even though both interactions remain present in the Hamiltonian. The suppression is therefore a direct coherent phase-cancellation effect, not a fitted proxy or an additional dephasing mechanism.

The branch state starts from Eq.~\eqref{eq:framework_initial_state} and evolves according to Eq.~\eqref{eq:framework_branch_evolution}, with \(\omega_i\) replaced by the total branch rate and with the same \(Q_{ij}\) defined in Eqs.~\eqref{eq:framework_Qij} and \eqref{eq:framework_branch_displacements}. The calculations use
\begin{equation}
    d=1,
    \qquad
    \frac{\Delta x}{d}=0.80,
    \qquad
    \Gamma_0=0.018,
    \qquad
    \frac{\ell_c}{d}=1.25.
    \label{eq:fig3_parameters}
\end{equation}

Panel (a) uses \(s=0.45\) and
\(\kappa_E\in\{0.20,1.00,3.00\}\). For panels (b)--(d), the electrostatic width ratio is fixed at \(\kappa_E=1.80\). This value is not claimed to be universal and represents a phenomenological profile broader than the gravitational mass profile. The scan covers
\begin{align}
    0.04\leq & s=\frac{\sigma_G}{d}\leq0.55,
    \qquad
    0.60\leq \tau\leq18,
\\ 
    -1.20\leq & \epsilon\leq5.20.
    \label{eq:fig3_scan}
\end{align}
The numerical grids are listed in Table~\ref{tab:numerical_grids}.

The Bell and teleportation diagnostics are evaluated without redefining them: \(\mathcal{B}_{\max}\) and \(F_{\rm tel}^{\rm opt}\) are obtained from Eqs.~\eqref{eq:framework_Tmn}--\eqref{eq:framework_ftel}. Their simultaneous thresholds are the same as in Fig.~\ref{fig:fig1}. To quantify the finite operational duration, we define
\begin{align}
    T_{\rm use}(s,\epsilon)
    =
    \int_{\tau_{\min}}^{\tau_{\max}}
    &\Theta\left[
    \mathcal{B}_{\max}(\tau,s,\epsilon)-2
    \right]\nonumber\\
    &\times\Theta\left[
    F_{\rm tel}^{\rm opt}(\tau,s,\epsilon)-\frac{2}{3}
    \right]
    d\tau,
    \label{eq:Tuse_definition}
\end{align}
where \(\Theta(x)\) is the Heaviside step function, \(\tau_{\min}=0.60\), and \(\tau_{\max}=18\). Thus, \(T_{\rm use}\) is the total dimensionless time interval during which the same state simultaneously violates the CHSH bound and beats the classical teleportation threshold.

\begin{figure*}[t]
    \centering
    \includegraphics[width=\linewidth]{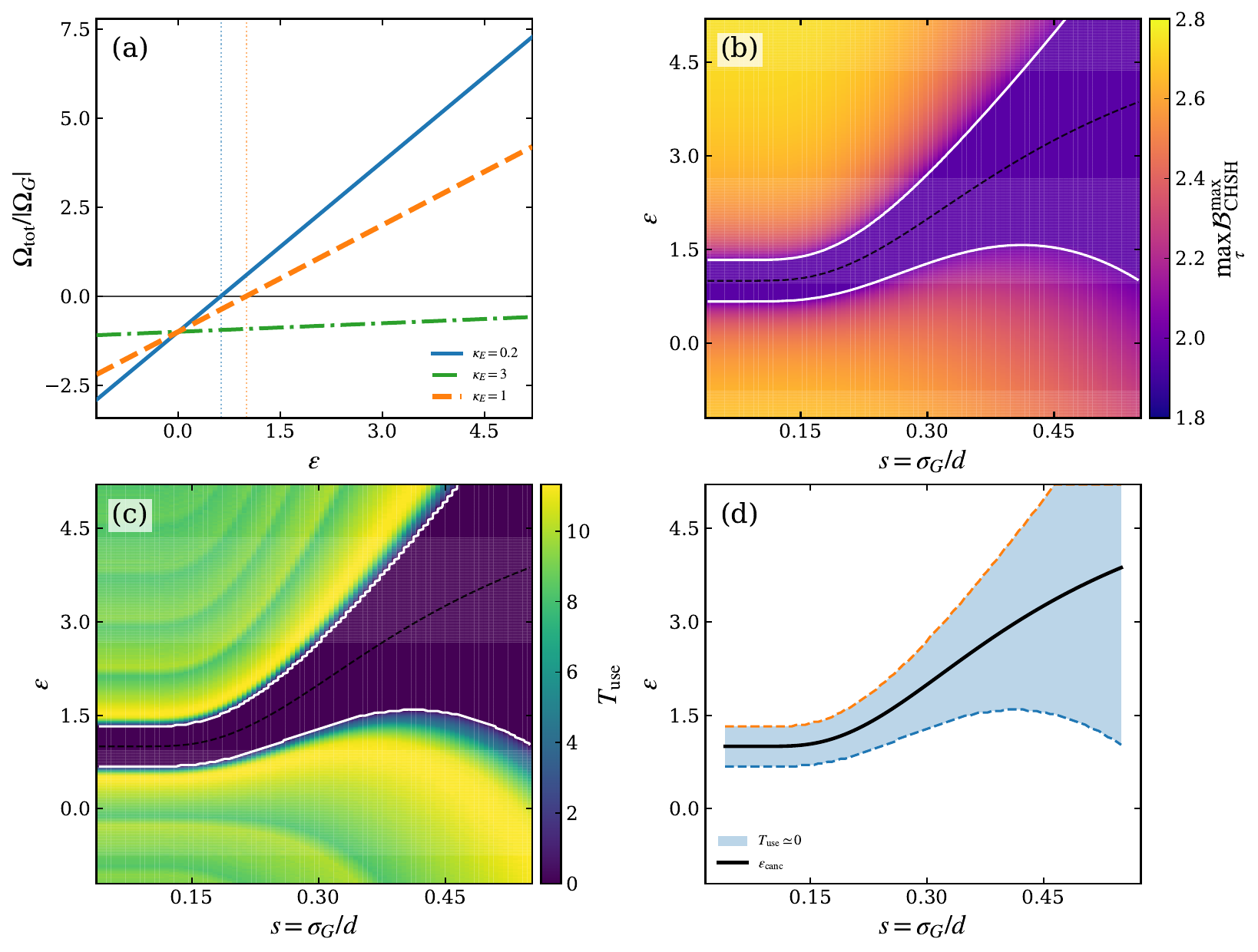}
    \caption{
Electrostatic-contamination tolerance of the finite-width gravitational cat channel.
The total branch-rate vector is \(\omega_{ab}^{\rm tot}=\omega_{ab}^{G}+\epsilon\omega_{ab}^{E}\), with \(\omega_{ab}^{G}=-dK(D_{ab},\sigma_G)\), \(\omega_{ab}^{E}=+dK(D_{ab},\sigma_E)\), and \(K(D,\sigma)=\operatorname{erf}[D/(2\sigma)]/D\). The code uses \(d=1\). Positive and negative \(\epsilon\) denote same-sign repulsive and opposite-sign effective contributions, respectively.
(a) Normalized total nonlocal phase rate \(\Omega_{\rm tot}/|\Omega_G|\) versus \(\epsilon\) at \(s=0.45\), for \(\kappa_E=0.20,1.00,3.00\). The zero crossing gives \(\epsilon_{\rm canc}=-\Omega_G/\Omega_E\).
(b) Maximum Bell--CHSH value versus \(s\) and \(\epsilon\) for \(\kappa_E=1.80\). White contours mark \(\mathcal{B}_{\rm CHSH}^{\max}=2\), and the black dashed curve marks \(\epsilon_{\rm canc}(s)\).
(c) Useful duration \(T_{\rm use}\), evaluated by the discrete sum in Eq.~\eqref{eq:Tuse_discrete}.
(d) Extracted cancellation band. The black curve is \(\epsilon_{\rm canc}(s)\), and the shaded region denotes \(T_{\rm use}\leq0.03\). The scans use \(0.04\leq s\leq0.55\), \(-1.20\leq\epsilon\leq5.20\), and \(0.60\leq\tau\leq18\), with \(d=1\), \(\Delta x/d=0.80\), \(\Gamma_0=0.018\), and \(\ell_c/d=1.25\).
}
    \label{fig:fig3}
\end{figure*}

In the code, this integral is evaluated on the uniform time grid as
\begin{equation}
    T_{\rm use}
    \simeq
    \Delta\tau
    \sum_k
    \mathbf{1}\!\left[\mathcal{B}_{\max}(\tau_k)>2\right]
    \mathbf{1}\!\left[F_{\rm tel}^{\rm opt}(\tau_k)>\frac{2}{3}\right].
    \label{eq:Tuse_discrete}
\end{equation}
Here, \(\Delta\tau\) is the uniform time-grid spacing, and \(\mathbf{1}[\mathcal C]\) is the indicator function, equal to one when the condition \(\mathcal C\) is satisfied and zero otherwise.
For the extracted ``dead band'' in Fig.~\ref{fig:fig3}(d), where the shaded region of suppressed operational usefulness is plotted there, the code uses \(T_{\rm use}\leq0.03\), which corresponds to the absence of a useful sampled time on the adopted grid.

Figure~\ref{fig:fig3} summarizes this electrostatic-contamination test. Panel (a) shows the normalized total nonlocal phase rate \(\Omega_{\rm tot}/|\Omega_G|\) as a function of the electrostatic amplitude \(\epsilon\), for a representative gravitational width \(s=\sigma_G/d=0.45\). Three values of \(\kappa_E\) are shown to illustrate how the finite-width profile of the electrostatic contribution changes the cancellation point and the slope of the total nonlocal phase. The zero crossing gives \(\epsilon_{\rm canc}\), where the electrostatic phase cancels the gravitational nonlocal phase.

Panel (b) shows the directly measurable quantity \(\max_{\tau}\mathcal{B}_{\rm CHSH}^{\max}\) as a function of \(s=\sigma_G/d\) and \(\epsilon\), for fixed \(\kappa_E=1.80\). The white contours mark the CHSH threshold \(\mathcal{B}_{\rm CHSH}^{\max}=2\), and the black dashed curve gives \(\epsilon_{\rm canc}(s)\). The suppression of Bell violation around this curve supports the interpretation that cancellation at the phase level translates into loss of operational nonlocality.

Panel (c) displays the useful duration \(T_{\rm use}\). This panel gives the most transparent operational summary: bright regions correspond to finite time intervals where both \(\mathcal{B}_{\rm CHSH}^{\max}>2\) and \(F_{\rm tel}^{\rm opt}>2/3\) hold simultaneously, while the dark band around \(\epsilon_{\rm canc}(s)\) identifies the strongly suppressed operational band. Thus, electrostatic contamination does not only shift the phase; it can eliminate the usable Bell--teleportation window.

Panel (d) extracts the cancellation/suppression-band structure from panel (c). The black curve gives the analytic cancellation point \(\epsilon_{\rm canc}(s)\), while the shaded band marks the region where \(T_{\rm use}\simeq0\). This panel provides a compact tolerance criterion: outside the shaded band, the finite-width gravitational channel retains a nonzero operational window; inside the band, electrostatic cancellation suppresses the simultaneous Bell and teleportation advantage.

The main finding of this figure is that residual electrostatic interactions do not simply add noise to the gravitational cat channel. Instead, because they contribute coherently to the same branch-dependent nonlocal phase, they can either reinforce or cancel the finite-width gravitational signal. The cancellation band in Fig.~\ref{fig:fig3} therefore gives a physically interpretable tolerance window for preserving operational gravitational-channel usefulness in the presence of electrostatic contamination.

\subsection{Gravity-amplitude estimation inside the useful window}
\label{subsec:communication_metrology}

We next ask whether the useful communication window also carries information about the strength of the gravitational interaction. This is a natural consistency test. A channel that is useful for Bell and teleportation tasks should also retain, at least in part, measurable information about the interaction that produced it. We therefore treat the dimensionless gravitational amplitude \(g\), multiplying all gravitational branch rates, as an unknown parameter and estimate how much Fisher information can be accessed through simple two-qubit Pauli-correlation measurements.

The total branch-rate model is
\begin{equation}
    \omega_{ab}^{\rm tot}(g)
    =
    g\,\omega_{ab}^{G}
    +
    \epsilon \omega_{ab}^{E},
    \qquad
    a,b\in\{L,R\},
    \label{eq:fig4_total_rate}
\end{equation}
where \(g\) is the gravitational amplitude to be estimated. The gravitational and electrostatic components are the finite-width rates in Eqs.~\eqref{eq:framework_grav_rate} and \eqref{eq:framework_elec_rate}.

\begin{figure*}[t]
    \centering
    \includegraphics[width=\linewidth]{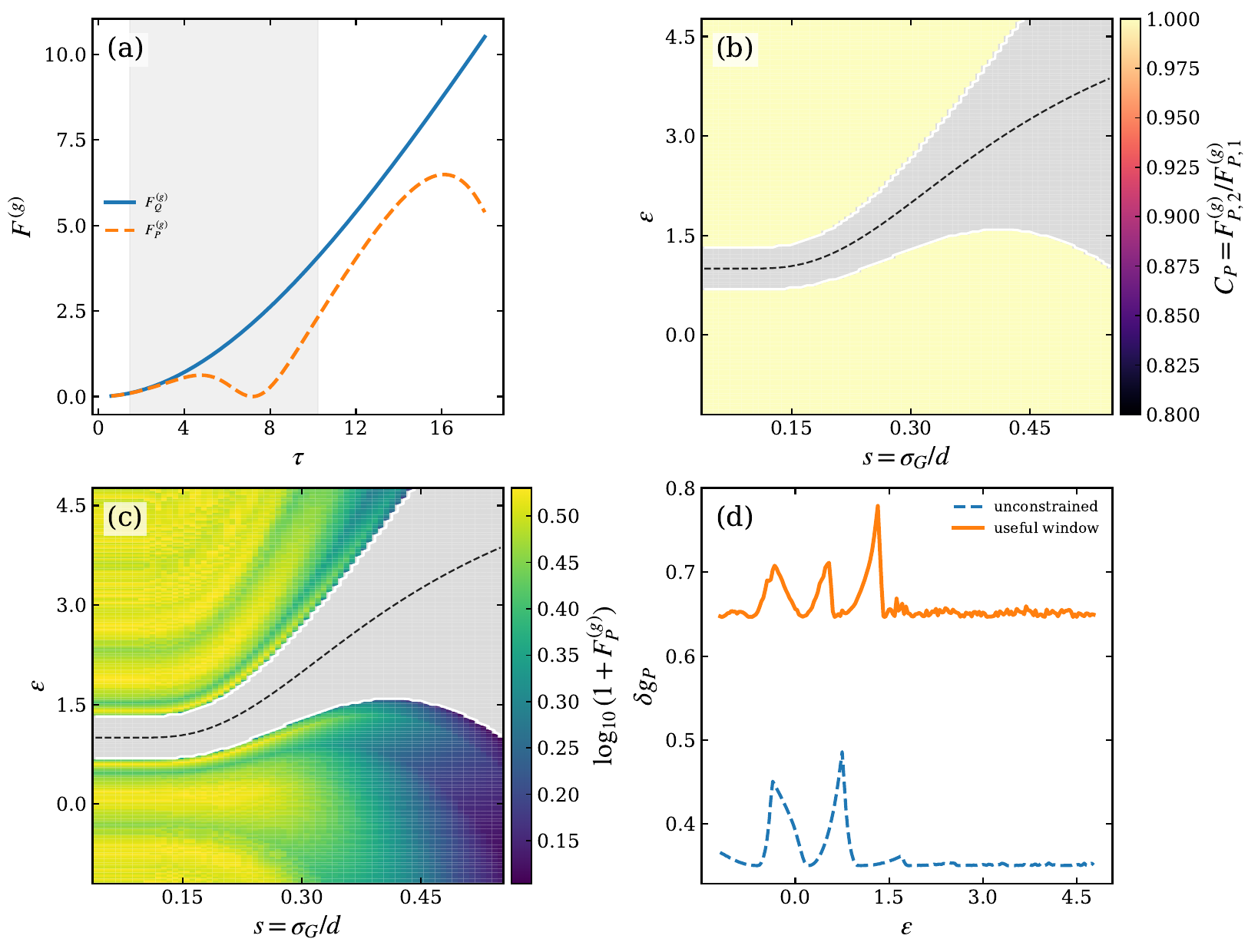}
    \caption{
Communication-compatible gravitational-amplitude estimation in the finite-width gravcat channel.
The total rate is \(\omega_{ab}^{\rm tot}=g\omega_{ab}^{G}+\epsilon\omega_{ab}^{E}\), evaluated at \(g=1\) with \(E_0\), \(\tau\), \(\Gamma_0\), and \(\epsilon\) held fixed under differentiation. The calculations use \(d=1\), \(\Delta x/d=0.80\), \(\Gamma_0=0.018\), \(\ell_c/d=1.25\), and \(\kappa_E=1.80\).
(a) Quantum and Pauli Fisher informations versus \(\tau\) at \(s_\ast=0.1908\) and \(\epsilon_\ast=-0.0007\). The shaded region satisfies both Bell and teleportation thresholds.
(b) Readout flexibility \(C_P=F_{P,2}^{(g)}/F_{P,1}^{(g)}\), evaluated at the same useful-window time that maximizes \(F_P^{(g)}\). Gray regions have no Bell--teleportation useful window.
(c) \(\log_{10}(1+F_P^{(g)})\), maximized over useful-window times. The white contour marks the useful-window boundary, and the dashed black curve marks \(\epsilon_{\rm canc}(s)\).
(d) Single-use Pauli-based Cramér--Rao scale versus \(\epsilon\). The dashed curve uses \(\max_{s,\tau}F_P^{(g)}\); the solid curve uses \(\max_{s,\tau\in\mathcal W_{\rm use}}F_P^{(g)}\).
}
    \label{fig:fig4}
\end{figure*}

The calculation is evaluated at \(g=1\). The reference energy \(E_0\), and hence the numerical variables \(\tau\), \(\Gamma_0\), and \(\epsilon\), are held fixed when the derivative with respect to \(g\) is taken. Figure~\ref{fig:fig4} uses \(\kappa_E=1.80\), \(d=1\), \(\Delta x/d=0.80\), \(\Gamma_0=0.018\), and \(\ell_c/d=1.25\). The corresponding numerical grids are listed in Table~\ref{tab:numerical_grids}.

For fixed \(g\), the state is propagated with Eq.~\eqref{eq:framework_branch_evolution} after replacing \(\omega_i\) by \(\omega_i^{\rm tot}(g)\). Since the dephasing factor is independent of \(g\) in the present phenomenological model, differentiating the density matrix acts only on the coherent gravitational phase:
\begin{equation}
    \partial_g \rho_{ij}(\tau;g)
    =
    -i
    \left(
    \omega_i^{G}-\omega_j^{G}
    \right)
    \tau\,
    \rho_{ij}(\tau;g).
    \label{eq:fig4_drho}
\end{equation}

The representative point used in panel (a) is selected at the sampled value of \(\epsilon\) closest to zero by maximizing the useful-window Pauli Fisher information over \(s\):
\begin{equation}
    s_\ast=0.1908,
    \qquad
    \epsilon_\ast=-0.0007.
    \label{eq:fig4_representative_point}
\end{equation}

The ultimate sensitivity is quantified by the quantum Fisher information \(F_Q^{(g)}\) defined in Eq.~\eqref{eq:framework_qfi}. The spectral sum includes only terms with \(\lambda_\mu+\lambda_\nu>0\), and it gives the best local estimation precision allowed by quantum measurement theory for the output state \(\rho(\tau;g)\).

To connect the metrological result to experimentally direct observables, we also compute the classical Fisher information associated with two-qubit Pauli-correlation measurements. For each Pauli product \(O_{mn}=\sigma_m\otimes\sigma_n\), with \(m,n\in\{x,y,z\}\), the measured correlation is \(T_{mn}=\operatorname{Tr}[\rho\,\sigma_m\otimes\sigma_n]\), the same Pauli-correlation matrix element already introduced in Eq.~\eqref{eq:framework_Tmn}. Since this measurement has binary outcomes, its classical Fisher information is
\begin{equation}
    F_{mn}^{(g)}
    =
    \frac{
    \left(
    \partial_g T_{mn}
    \right)^2
    }
    {
    1-T_{mn}^2}.
    \label{eq:fig4_pauli_fisher_single}
\end{equation}
The experimentally accessible Pauli Fisher information is
\begin{equation}
    F_P^{(g)}
    =
    \max_{m,n\in\{x,y,z\}}
    F_{mn}^{(g)}.
    \label{eq:fig4_pauli_fisher}
\end{equation}
This is the classical Fisher information of a directly measurable Pauli-correlation readout, rather than a proxy constructed from an unrelated observable.

We restrict the communication-compatible metrological region to the useful-window set
\begin{equation}
    \mathcal{W}_{\rm use}
    =
    \left\{
    (s,\tau):
    \mathcal{B}_{\max}(\tau,s)>2,\
    F_{\rm tel}^{\rm opt}(\tau,s)>\frac{2}{3}
    \right\},
    \label{eq:Wuse_definition}
\end{equation}
i.e.\ the points satisfying the Bell and teleportation thresholds in Eqs.~\eqref{eq:framework_chsh} and \eqref{eq:framework_ftel} at the fixed \(\epsilon\). Only in this region do we report the communication-compatible Pauli Fisher information.

For every \((s,\epsilon)\), the code first restricts the time grid to this useful window and then selects the time that maximizes \(F_P^{(g)}\). The largest and second-largest Pauli Fisher informations are evaluated at that same selected time, so the flexibility ratio below compares two readouts for one common output state.

To quantify readout flexibility, let
\begin{equation}
    F_{P,1}^{(g)}
    =
    \max_{m,n}F_{mn}^{(g)}
\end{equation}
and let \(F_{P,2}^{(g)}\) be the second-largest value among the nine Pauli-correlation Fisher informations. We define
\begin{equation}
    C_P
    =
    \frac{
    F_{P,2}^{(g)}
    }{
    F_{P,1}^{(g)}
    },
    \label{eq:fig4_readout_flexibility}
\end{equation}
where the subscript \(P\) labels ``Pauli''; \(C_P\) is a readout-flexibility ratio.  Values \(C_P\simeq1\) mean that the estimation protocol is not tied to a single finely tuned Pauli setting.

The single-use Pauli-based Cramér--Rao scale plotted in Fig.~\ref{fig:fig4} is
\begin{equation}
    \delta g_P^{(1)}
    =
    \frac{1}{\sqrt{F_P^{(g)}}}.
    \label{eq:fig4_cramer_rao}
\end{equation}
For \(N\) statistically independent repetitions, the bound becomes \(\delta g_P\geq1/\sqrt{N F_P^{(g)}}\). We compare the unconstrained single-use scale with the communication-compatible scale obtained only inside the Bell--teleportation useful window.

Figure~\ref{fig:fig4} summarizes the metrological behavior. Panel (a) compares the quantum Fisher information \(F_Q^{(g)}\) with the experimentally accessible Pauli Fisher information \(F_P^{(g)}\) at a representative operating point. The shaded region marks the time interval where the same state is Bell-nonlocal and teleportation-useful. This panel shows that the finite-width gravcat channel carries measurable information about the gravitational amplitude, and that a Pauli-correlation readout captures a substantial part of the available sensitivity during the operational window.

Panel (b) shows the readout-flexibility ratio \(C_P=F_{P,2}^{(g)}/F_{P,1}^{(g)}\). The bright values close to unity indicate that the best and second-best Pauli readouts are nearly equivalent over most of the useful region. This is experimentally favorable: the metrological signal is not fragile against a small change in the chosen Pauli-correlation measurement. The grey region denotes parameters where no Bell--teleportation useful window exists, so no communication-compatible metrological claim is made there.

Panel (c) gives the magnitude of the accessible Pauli Fisher information inside the useful window, plotted as \(\log_{10}(1+F_P^{(g)})\). The white contour marks the boundary of the useful region, while the black dashed curve marks the electrostatic cancellation point \(\epsilon_{\rm canc}(s)\). The Fisher information is suppressed near the cancellation band, showing that the same electrostatic mechanism that suppresses Bell and teleportation usefulness also weakens gravitational-amplitude estimation.

Panel (d) converts the Fisher information into the direct Cramér--Rao precision \(\delta g_P\). At each \(\epsilon\), both curves are additionally optimized over the scanned widths. The unconstrained curve uses
\[
\max_{s,\tau}F_P^{(g)},
\]
whereas the useful-window curve uses
\[
\max_{s,\tau\in\mathcal{W}_{\rm use}}F_P^{(g)}.
\]
The useful-window precision is therefore more conservative, but it has the stronger interpretation that estimation is performed only where the channel remains Bell-nonlocal and teleportation-useful.

The main observation from Fig.~\ref{fig:fig4} is that the finite-width gravcat channel has a common operational window for communication and metrology. In this window, the same noisy channel can simultaneously support Bell violation, teleportation advantage, and gravitational-amplitude estimation using directly measurable Pauli correlations. Moreover, the near-unity values of \(C_P\) indicate that the Pauli readout is experimentally flexible rather than dependent on a single fine-tuned measurement setting.

\subsection{Robustness against dephasing}
\label{subsec:global_robustness_scaling}

We now ask a global robustness question: how much spatially correlated dephasing can the finite-width gravcat channel tolerate before it loses its joint operational usefulness? In the previous sections, the channel was tested through Bell nonlocality, teleportation advantage, electrostatic cancellation, and gravitational-amplitude estimation. Here we combine these ideas into a single robustness criterion and determine the critical dephasing strength beyond which the operational window disappears.

For each wave-packet width \(s=\sigma_G/d\) and electrostatic amplitude ratio \(\epsilon\), we define the critical dephasing strength \(\Gamma_c(s,\epsilon)\) as the largest sampled value of \(\Gamma_0\) for which the useful duration exceeds a fixed threshold:
\begin{equation}
    \Gamma_c(s,\epsilon)
    =
    \max
    \left\{
    \Gamma_0:
    T_{\rm use}(s,\epsilon,\Gamma_0)>T_{\rm min}
    \right\}.
    \label{eq:critical_dephasing_definition}
\end{equation}
The code uses 46 equally spaced dephasing values over \(0\leq\Gamma_0\leq0.16\), 120 time points over \(0.60\leq\tau\leq18\), and
\begin{equation}
    T_{\rm min}
    =
    \max\{0.10,1.5\Delta\tau\}
    =
    0.219.
    \label{eq:Tmin_value}
\end{equation}
Thus, \(\Gamma_c\) is a grid-resolved numerical estimate. If no sampled dephasing value satisfies the criterion, the code stores that parameter point as undefined rather than assigning \(\Gamma_c=0\).

The electrostatic cancellation point is the same phase-level condition defined in Eqs.~\eqref{eq:framework_total_omega} and \eqref{eq:framework_eps_cancel}. Near this line, the gravitational and residual electrostatic nonlocal phases cancel, so the Bell--teleportation window is expected to be fragile. The key question is whether the strong reduction of \(\Gamma_c\) near \(\epsilon_{\rm canc}\) follows a systematic trend.

To test this, we introduce the distance from the cancellation line,
\begin{equation}
    \Delta\epsilon
    =
    \left|
    \epsilon-\epsilon_{\rm canc}(s)
    \right|,
    \label{eq:delta_epsilon_definition}
\end{equation}
and bin the computed values of \(\Gamma_c(s,\epsilon)\) according to \(\Delta\epsilon\). The binned median and interquartile range (IQR) are then fitted in the near-cancellation regime using the power-law form
\begin{equation}
    \Gamma_c
    \simeq
    A
    \left|
    \epsilon-\epsilon_{\rm canc}
    \right|^\nu,
    \label{eq:critical_dephasing_scaling}
\end{equation}
where \(A\) is the fitted prefactor and \(\nu\) is the fitted scaling exponent. The quantity \(R^2\) reported below is the coefficient of determination of the fit.

\begin{figure*}[t]
    \centering
    \includegraphics[width=\linewidth]{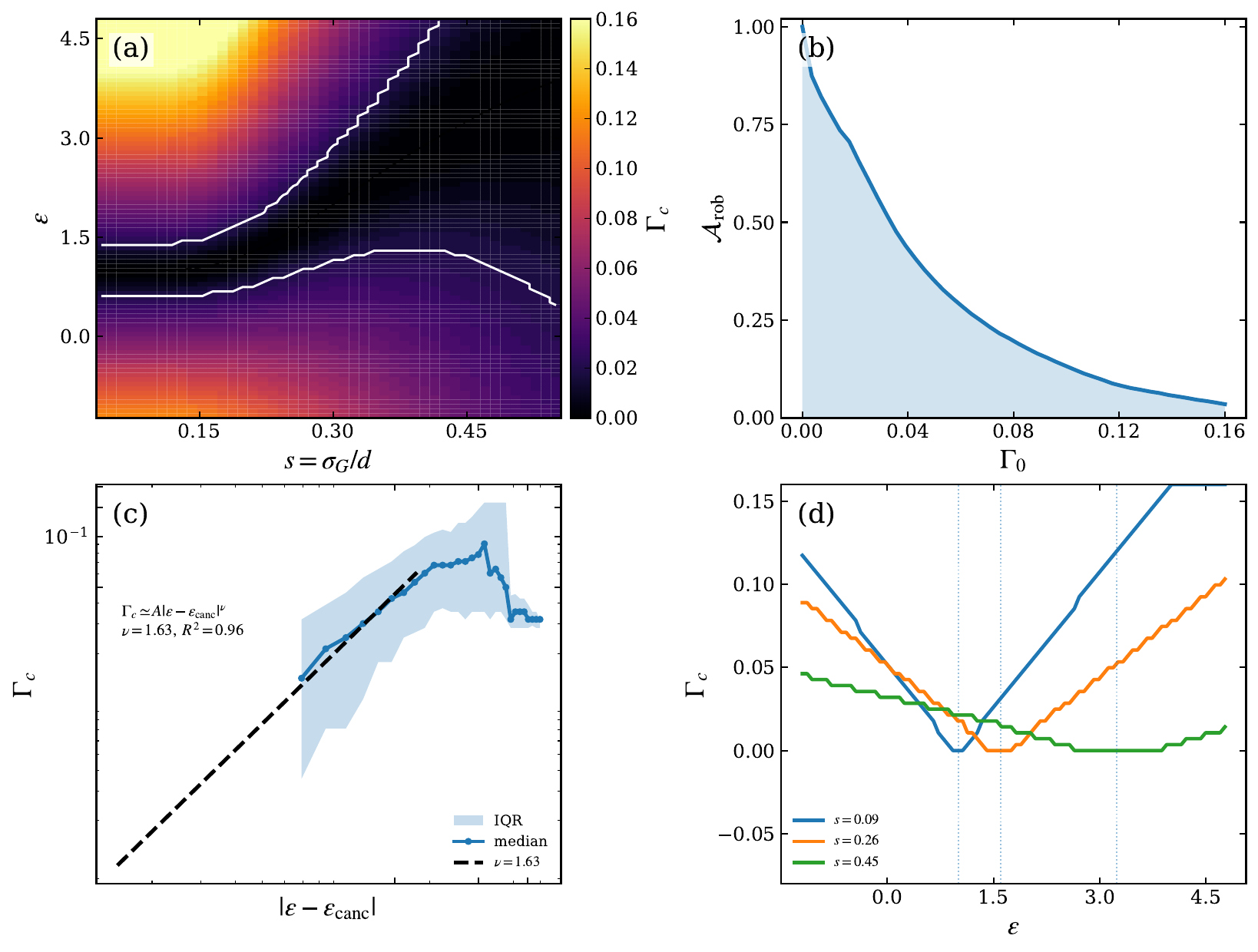}
    \caption{
Global robustness and critical-dephasing scaling of the finite-width gravcat operational phase. The calculations use \(d=1\), \(\Delta x/d=0.80\), \(\ell_c/d=1.25\), and \(\kappa_E=1.80\).
(a) Grid-resolved critical dephasing \(\Gamma_c(s,\epsilon)\), defined as the largest sampled \(\Gamma_0\) satisfying \(T_{\rm use}>T_{\min}=0.219\). Undefined points have no accepted sampled dephasing value. The dashed curve marks \(\epsilon_{\rm canc}(s)\), and the white contour is \(\Gamma_c=0.02\).
(b) Robust fraction \(\mathcal A_{\rm rob}\) over the finite-\(\Gamma_c\) set \(\mathcal V\).
(c) Binned-median scaling versus \(|\epsilon-\epsilon_{\rm canc}|\). The fit over \(0.15<|\epsilon-\epsilon_{\rm canc}|<1.80\) gives \(A=0.023519\), \(\nu=1.625282\), and \(R^2=0.963324\).
(d) Cuts at \(s=0.09\), \(0.26\), and \(0.45\).
}
    \label{fig:fig5}
\end{figure*}

For the parameter window used in Fig.~\ref{fig:fig5}, the finite, positive data are grouped into 33 bins in \(\Delta\epsilon\), and the medians are fitted over
\[
0.15<\Delta\epsilon<1.80.
\]
The fit gives
\begin{equation}
    A=0.023519,
    \qquad
    \nu=1.625282,
    \qquad
    R^2=0.963324.
    \label{eq:fig5_exponent_result}
\end{equation}
This is a numerical scaling trend for the present model, grid, fitting window, and acceptance threshold, not a universal critical exponent.

Figure~\ref{fig:fig5} uses the grids listed in Table~\ref{tab:numerical_grids}. Panel (a) shows the critical dephasing map \(\Gamma_c(s,\epsilon)\), and its white contour is \(\Gamma_c=0.02\).
 Bright regions indicate parameter regimes where the channel remains operationally useful even under stronger dephasing, while dark regions indicate rapid loss of Bell violation and teleportation advantage. The dashed curve marks \(\epsilon_{\rm canc}(s)\), and the suppression of \(\Gamma_c\) around this curve supports the interpretation that electrostatic cancellation strongly shapes the robustness of the channel.

Panel (b) compresses the two-dimensional phase diagram into a robust fraction over the valid set
\begin{equation}
    \mathcal{V}
    =
    \left\{
    (s,\epsilon):
    \Gamma_c(s,\epsilon)\ \text{is finite}
    \right\}.
    \label{eq:robust_valid_set}
\end{equation}
The quantity plotted by the code is
\begin{equation}
    \mathcal{A}_{\rm rob}(\Gamma_0)
    =
    \frac{
    \#\left\{
    (s,\epsilon)\in\mathcal{V}:
    \Gamma_c(s,\epsilon)\geq\Gamma_0
    \right\}
    }{
    \#\mathcal{V}
    }.
    \label{eq:robust_area_fraction}
\end{equation}
It therefore measures the fraction of parameter points possessing a finite critical-dephasing estimate that remain useful at the specified \(\Gamma_0\).

Panel (c) gives the cancellation scaling. The horizontal axis is \(|\epsilon-\epsilon_{\rm canc}|\), and the vertical axis is \(\Gamma_c\). The median trend and interquartile band (or IQR) show how the critical dephasing increases as one moves away from the cancellation line. The dashed fit gives the scaling exponent \(\nu\simeq1.63\), indicating an approximate power-law recovery of noise tolerance away from electrostatic cancellation.

Panel (d) shows the cuts at \(s=0.09\), \(0.26\), and \(0.45\). Each curve exhibits a strong dip near its corresponding cancellation point, while the tolerated dephasing away from cancellation depends on \(s\).

The main observation from Fig.~\ref{fig:fig5} is that the finite-width gravcat channel supports a finite noise-tolerant operational phase, but this phase is reduced in a controlled way near electrostatic cancellation. The fitted scaling in Eq.~\eqref{eq:critical_dephasing_scaling} provides a compact quantitative summary of this reduction. In this sense, the robustness of Bell--teleportation usefulness is governed by the distance from the electrostatic-cancellation line.

\subsection{Distinguishing gravity from a calibrated electrostatic mimic}
\label{subsec:geometry_calibrated_discrimination}

We then test a more subtle possibility: a residual electrostatic interaction may imitate the gravitational channel after being calibrated at one reference geometry. This is a stronger test than comparing gravity with an unmatched electrostatic model. If the electrostatic model were allowed to be re-tuned at every wave-packet width, the comparison would be too forgiving and could hide the different finite-width scaling. We therefore use a fixed calibration protocol: the electrostatic mimic is matched once at a reference width \(s_{\rm cal}\), and the same calibrated mimic is then tested at other widths.

The gravitational rates are the finite-width rates of Eq.~\eqref{eq:framework_grav_rate}, with \(s=\sigma_G/d\). The raw electrostatic rates use Eq.~\eqref{eq:framework_elec_rate} with \(\sigma_E=\kappa_E\sigma_G\). For \(X=G\) and \(X=E\), the nonlocal rates are obtained from Eqs.~\eqref{eq:framework_nonlocal_rate} and \eqref{eq:framework_nonlocal_rate_E}, respectively. This common definition is essential: the calibration compares the part of each interaction that cannot be removed by local rephasings, rather than comparing arbitrary branch energies.

The electrostatic mimic is calibrated at the reference width
\(s_{\rm cal}=0.25\). Here
\begin{equation}
    \Omega_E^{0}(s)
    =
    \omega_{LL}^{E,0}
    +
    \omega_{RR}^{E,0}
    -
    \omega_{LR}^{E,0}
    -
    \omega_{RL}^{E,0}
    \label{eq:fig6_raw_electrostatic_rate}
\end{equation}
is the invariant rate of the uncalibrated positive-sign electrostatic kernel. A single scale factor \(\alpha_E\) is chosen such that
\begin{equation}
    \Omega_G(s_{\rm cal})
    =
    \Omega_E^{\rm cal}(s_{\rm cal})
    =
    \alpha_E
    \Omega_E^{0}(s_{\rm cal}).
    \label{eq:fig6_calibration_condition}
\end{equation}

\begin{figure*}[t]
    \centering
    \includegraphics[width=\linewidth]{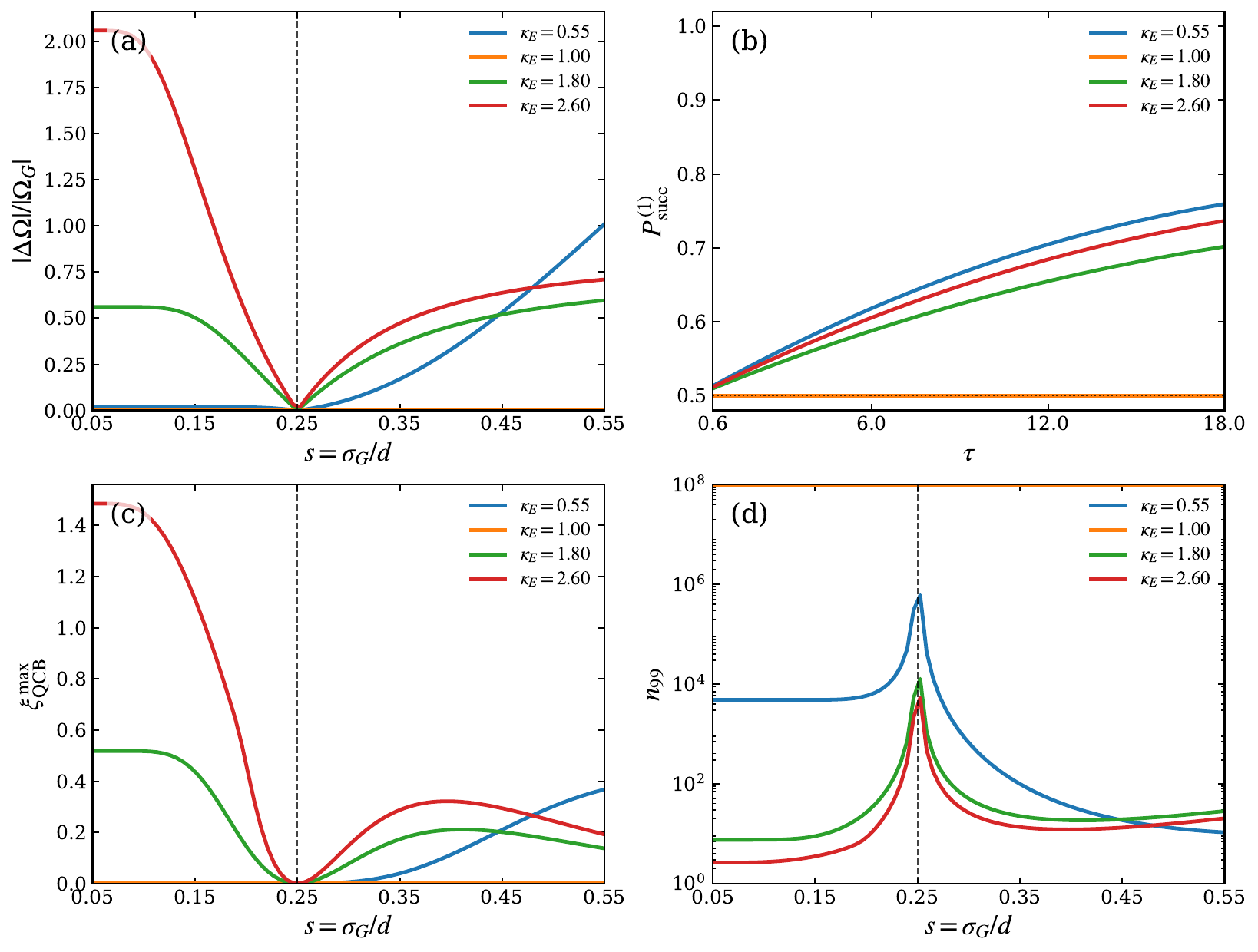}
    \caption{
Geometry-calibrated gravity--electrostatic benchmark discrimination.
The effective electrostatic mimic is calibrated at \(s_{\rm cal}=0.25\) and then tested without re-fitting. The calculations use \(d=1\), \(\Delta x/d=0.80\), \(\Gamma_0=0.018\), and \(\ell_c/d=1.25\), with \(\kappa_E=0.55,1.00,1.80,2.60\) and the corresponding negative calibration factors listed in Eq.~\eqref{eq:fig6_calibration_values}.
(a) Normalized phase mismatch \(|\Delta\Omega|/|\Omega_G|\). The dashed line marks \(s_{\rm cal}=0.25\).
(b) Single-shot success probability versus \(\tau\) at \(s_{\rm test}=0.50\). The dotted line marks random guessing.
(c) Time-window maximum \(\xi_{\rm QCB}^{\max}\) over \(0.60\leq\tau\leq18\), computed on the uniform \(q\)-grid specified in Table~\ref{tab:numerical_grids}.
(d) Real-valued asymptotic estimate \(n_{99}^{\rm est}\) for \(P_{\rm tar}=0.01\), plotted with a cap of \(10^8\). The \(\kappa_E=1\) profile remains exactly indistinguishable, while \(\kappa_E\neq1\) produces discrimination through profile mismatch.
}
    \label{fig:fig6}
\end{figure*}

Therefore,
\begin{equation}
    \alpha_E
    =
    \frac{
    \Omega_G(s_{\rm cal})
    }{
    \Omega_E^{0}(s_{\rm cal})
    }.
    \label{eq:fig6_alpha_calibration}
\end{equation}
The same \(\alpha_E\) is then held fixed for all other values of \(s\). This gives the calibrated electrostatic nonlocal rate
\begin{equation}
    \Omega_E^{\rm cal}(s)
    =
    \alpha_E\Omega_E^{0}(s).
    \label{eq:fig6_calibrated_elec_rate}
\end{equation}
The code uses
\begin{equation}
\begin{array}{c|cccc}
\kappa_E
&0.55&1.00&1.80&2.60\\
\hline
\alpha_E
&-0.979696&-1.000000&-1.560934&-3.057872
\end{array}
\label{eq:fig6_calibration_values}
\end{equation}
The negative factors are required because the raw electrostatic kernel is defined with the sign opposite to gravity. Thus, Fig.~\ref{fig:fig6} uses a separately calibrated opposite-sign effective electrostatic mimic, not the positive-\(\epsilon\) repulsive contaminant used in Figs.~\ref{fig:fig3}--\ref{fig:fig5}.

To avoid falsely distinguishing the two models through removable local phases, we compare only the nonlocal branch-phase component. Equivalently, after computing \(\Omega_X(s)\), we use the canonical nonlocal branch-rate vector
\begin{equation}
    \boldsymbol{\omega}_X^{\rm nl}(s)
    =
    \left(
    \frac{\Omega_X(s)}{4},
    -\frac{\Omega_X(s)}{4},
    -\frac{\Omega_X(s)}{4},
    \frac{\Omega_X(s)}{4}
    \right),
    \label{eq:fig6_canonical_rates}
\end{equation}
in the ordered basis \(\{LL,LR,RL,RR\}\). This makes the discrimination conservative: any remaining difference comes from the finite-width scaling of the nonlocal interaction rather than from local phase offsets.

The corresponding noisy branch states are
{\small \begin{equation}
    \rho_{ij}^{X}(\tau;s)
    =
    \frac{1}{4}
    \exp\left[
    -i
    \left(
    \omega_i^{X,{\rm nl}}(s)
    -
    \omega_j^{X,{\rm nl}}(s)
    \right)
    \tau
    \right]
    \exp\left[
    -\Gamma_0 Q_{ij}\tau
    \right],
    \label{eq:fig6_channel_state}
\end{equation}}
where \(X=G,E_{\rm cal}\), \(Q_{ij}\) is the same spatially correlated dephasing matrix used in the previous figures, and \(\tau=tE_0/\hbar\).

The single-shot distinguishability is evaluated with the trace distance and equal-prior success probability defined in Eqs.~\eqref{eq:framework_trace_distance} and the following equation in Sec.~\ref{subsec:framework_metrology_discrimination}. In the present subsection these quantities are evaluated for \(\rho_G(\tau;s)\) and the calibrated mimic \(\rho_E^{\rm cal}(\tau;s)\). A success probability of \(1/2\) is random guessing, while values above \(1/2\) provide a single-use discrimination advantage.

For many-copy discrimination, we use the quantum Chernoff exponent in Eq.~\eqref{eq:framework_qcb}, again with \(\rho_E\) interpreted as the calibrated electrostatic state. The largest many-copy discrimination exponent at each width is then
\begin{equation}
    \xi_{\rm QCB}^{\max}(s,\kappa_E)
    =
    \max_{\tau\in[0.60,18]}
    \xi_{\rm QCB}(\tau;s,\kappa_E).
    \label{eq:fig6_qcb_max}
\end{equation}

Using the asymptotic Chernoff error scaling stated after Eq.~\eqref{eq:framework_qcb}, the code reports the real-valued estimate of the number of copies needed to bring the discrimination error probability down to a chosen target level \(P_{\rm tar}\),
\begin{equation}
    n_{\rm tar}^{\rm est}
    =
    \frac{
    \log[1/(2P_{\rm tar})]
    }{
    \xi_{\rm QCB}^{\max}
    }.
    \label{eq:fig6_required_copies}
\end{equation}
In Fig.~\ref{fig:fig6}, \(P_{\rm tar}=0.01\), and the corresponding quantity is denoted by \(n_{99}^{\rm est}\). It is an asymptotic copy-complexity estimate rather than an integer run count. The code assigns \(n_{99}^{\rm est}=\infty\) when \(\xi_{\rm QCB}^{\max}=0\), and caps the plotted values at \(10^8\).

Figure~\ref{fig:fig6} shows the result of this geometry-calibrated discrimination test. Panel (a) plots the normalized finite-width mismatch with \(\Delta\Omega(s)\equiv\Omega_E^{\rm cal}(s)-\Omega_G(s)\),
\begin{equation}
    \frac{|\Delta\Omega|}{|\Omega_G|}
    =
    \frac{
    |\Omega_E^{\rm cal}(s)-\Omega_G(s)|
    }{
    |\Omega_G(s)|
    }.
    \label{eq:fig6_mismatch}
\end{equation}
All curves vanish at \(s=s_{\rm cal}\), as required by the calibration condition. Away from \(s_{\rm cal}\), the mismatch grows. This shows that a single electrostatic calibration can match the gravitational nonlocal phase at one geometry, but it does not reproduce the finite-width gravitational scaling over the full range of \(s\).

Panel (b) gives the single-shot success probability \(P_{\rm succ}^{(1)}\) at the representative test width \(s_{\rm test}=0.50\). The curve with \(\kappa_E=1\) remains at \(P_{\rm succ}^{(1)}=1/2\), which is the expected null case because the electrostatic and gravitational finite-width profiles are identical after calibration when \(\sigma_E=\sigma_G\). For \(\kappa_E\neq1\), the success probability rises above \(1/2\), showing that the calibrated electrostatic mimic can be rejected with a finite single-shot advantage.

Panel (c) shows the optimized quantum Chernoff exponent \(\xi_{\rm QCB}^{\max}\). It vanishes at the calibration point and becomes nonzero away from it. This demonstrates that the discrimination advantage is not only a single-shot trace-distance effect, but also appears in the many-copy asymptotic error exponent.

Panel (d) converts the Chernoff exponent into \(n_{99}^{\rm est}\). At \(s_{\rm test}=0.50\), the code obtains
\[
P_{\rm succ}^{(1),\max}
=
0.7595,\ 0.5000,\ 0.7019,\ 0.7367
\]
and
\[
n_{99}^{\rm est}
=
12.96,\ \infty,\ 22.79,\ 16.01
\]
for \(\kappa_E=0.55,1.00,1.80,2.60\), respectively. The nontrivial maxima occur at the upper boundary \(\tau=18\) and are therefore maxima only within the adopted time window.

The calibrated mimic is exactly indistinguishable from gravity for \(\kappa_E=1\) at every tested width because the two finite-width profiles then coincide. Thus, varying the wave-packet width provides discrimination only when the electrostatic and gravitational profiles scale differently, \(\kappa_E\neq1\), and only against the calibrated benchmark family considered here.

\subsection{When entanglement survives but usefulness fails}
\label{subsec:resource_fracture}

The previous sections indicate that the finite-width gravitational cat channel can lose Bell--teleportation usefulness near the electrostatic cancellation line. This loss, however, is not the same as a complete disappearance of quantum correlations. There is a parameter regime in which ordinary entanglement remains finite while the channel fails to satisfy the stricter operational requirements for nonlocality, teleportation, and quantum transmission. We refer to this separation as resource fracture.

For each output state \(\rho_{AB}(\tau;s,\epsilon)\), we evaluate the concurrence, the maximal Bell--CHSH parameter, the optimal teleportation fidelity, and the coherent information using Eqs.~\eqref{eq:framework_concurrence}, \eqref{eq:framework_chsh}, \eqref{eq:framework_ftel}, and \eqref{eq:framework_Ic}. Positive coherent information is used as a transmission-oriented lower-bound indicator, not as a necessary condition for every possible quantum task.

The strict same-time operational margin \(\mathcal{O}(\tau;s,\epsilon)\), its finite-time maximum \(\mathcal{O}_{\max}(s,\epsilon)\), and the maximum concurrence \(C_{\max}(s,\epsilon)\) are already defined in Eqs.~\eqref{eq:framework_operational_margin}, \eqref{eq:framework_Omax}, and \eqref{eq:framework_Cmax}. The common optimization interval is \(0.60\leq\tau\leq18\). Therefore, \(\mathcal{O}_{\max}>0\) means that Bell violation, teleportation advantage, and positive coherent information occur for the same output state at some finite time. The condition \(\mathcal{O}_{\max}\leq0\) means that this complete joint criterion fails throughout the scanned interval.

The three regions \(\mathsf U\), \(\mathsf F\), and \(\mathsf D\) are defined once in Eqs.~\eqref{eq:framework_U}--\eqref{eq:framework_D}. In particular, the numerical fractured regime satisfies \(C_{\max}>C_{\min}=10^{-4}\) while \(\mathcal{O}_{\max}\leq0\): the state remains entangled above the adopted tolerance, but the stronger operational channel condition has already failed. This separation is the central point of the classification and does not require a second set of equations in the Results section.

\begin{figure*}[t]
    \centering
    \includegraphics[width=\linewidth]{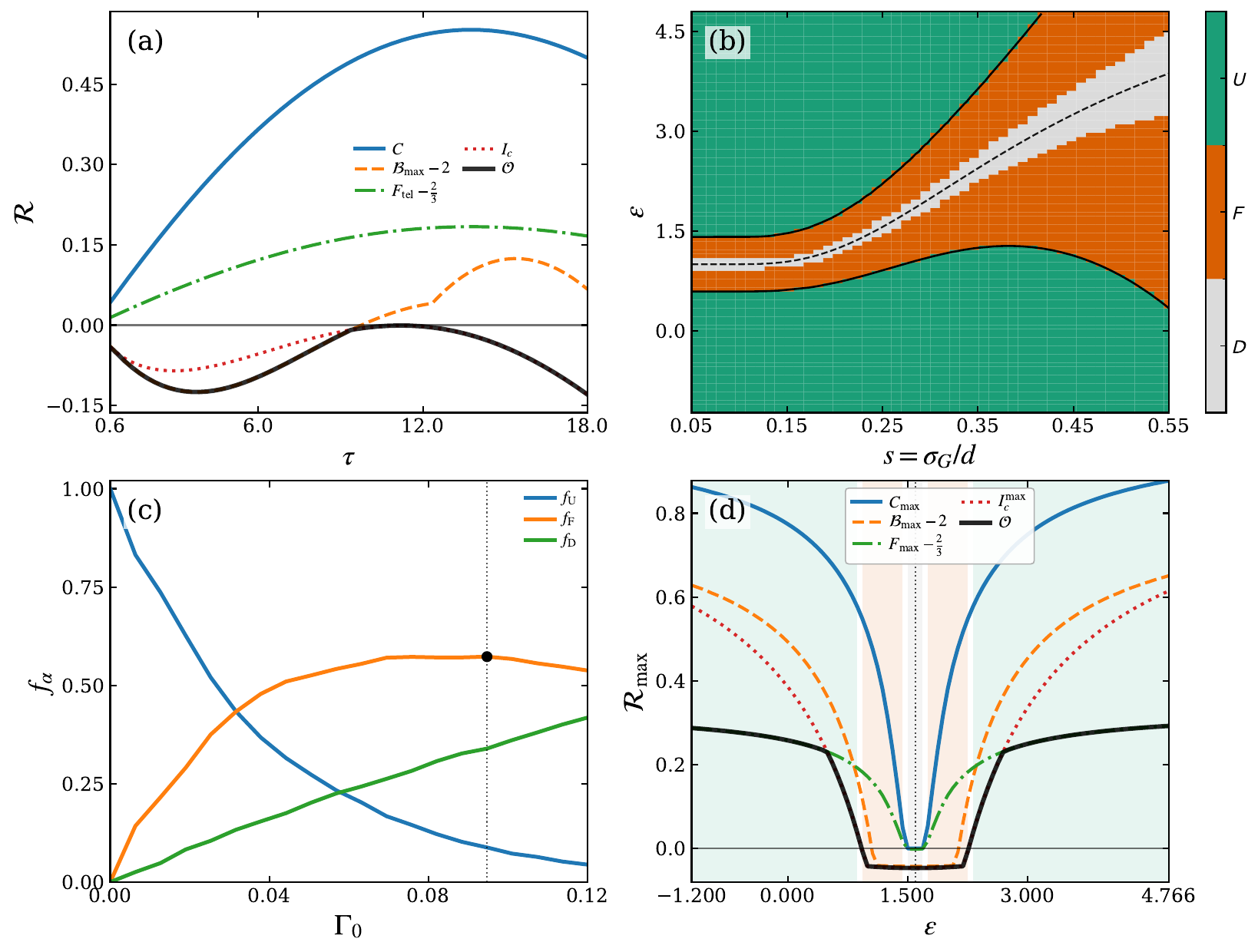}
    \caption{
Resource fracture of the finite-width gravitational cat channel. The parameters are \(d=1\), \(\Delta x/d=0.80\), \(\ell_c/d=1.25\), \(\kappa_E=1.80\), and \(C_{\min}=10^{-4}\).
(a) Time trace at \(s_\ast=0.30510\), \(\epsilon_\ast=2.94508\), and \(\Gamma_0=0.018\). The concurrence remains positive above tolerance while the joint operational margin stays non-positive.
(b) Categorical map in the \((s,\epsilon)\) plane. The regions \(\mathsf U\), \(\mathsf F\), and \(\mathsf D\) are defined by Eqs.~\eqref{eq:framework_U}--\eqref{eq:framework_D}. The dashed curve is \(\epsilon_{\rm canc}\), and the solid curve marks \(\mathcal O_{\max}=0\).
(c) Area fractions versus \(\Gamma_0\), calculated on the separate coarser grid listed in Table~\ref{tab:numerical_grids}. The fractured fraction peaks at \(0.572917\) for \(\Gamma_0=0.094737\).
(d) Cut at \(s=0.25\), showing that \(C_{\max}>C_{\min}\) can coexist with \(\mathcal O_{\max}\leq0\).
}
\label{fig:resource_fracture}
\end{figure*}

Figure~\ref{fig:resource_fracture} shows the resulting resource-fracture structure. Panel (a) uses the representative point (chosen, by inspection of panel (b), to lie inside the fractured region \(\mathsf{F}\), so as to illustrate the separation between entanglement survival and operational usefulness that is the subject of this subsection):
\begin{equation}
    s_\ast=0.30510,
    \qquad
    \epsilon_\ast=2.94508,
    \qquad
    \Gamma_0=0.018.
    \label{eq:fig7_representative_point}
\end{equation}
At this point,
\begin{align}
    &\max_\tau C =0.551867,\nonumber\\
&\max_\tau(\mathcal B_{\max}-2)=0.124436,\nonumber\\
    &\max_\tau(F_{\rm tel}^{\rm opt}-2/3)=0.183956,\\
    &\max_\tau I_c=-0.000632,
\nonumber\\
    &\mathcal O_{\max}=-0.000632.\nonumber
\label{eq:fig7_representative_values}
\end{align}
Thus, the individual Bell and teleportation thresholds can be crossed at some times, but the same-time criterion remains non-positive because coherent information never becomes positive.

Panel (b) shows the corresponding phase map in the \((s,\epsilon)\) plane. The fully operational region \(\mathsf{U}\), the fractured region \(\mathsf{F}\), and the depleted region \(\mathsf{D}\) appear as distinct domains. The dashed curve marks the electrostatic cancellation line \(\epsilon_{\rm canc}\), while the solid boundary marks the transition \(\mathcal{O}_{\max}=0\). The fractured region forms an intermediate domain between the operational and depleted regimes, showing that the loss of channel usefulness does not coincide with the immediate loss of entanglement.

Panel (c) uses the separate coarser area-fraction grid listed in Table~\ref{tab:numerical_grids}, on which we evaluate, for each region \(X\in\{\mathsf U,\mathsf F,\mathsf D\}\), the sampled area fraction
\begin{equation}
    f_X(\Gamma_0)
    =
    \frac{
    \#\{(s,\epsilon)\ \text{on the coarser grid}: (s,\epsilon)\in X\}
    }{
    \#\{(s,\epsilon)\ \text{on the coarser grid}\}
    },
    \label{eq:fXdefinition}
\end{equation}
so that \(f_{\mathsf U}+f_{\mathsf F}+f_{\mathsf D}=1\) at every sampled \(\Gamma_0\). At \(\Gamma_0=0.018\), the code gives
\[
f_{\mathsf U}=0.626953,
\qquad
f_{\mathsf F}=0.290365,
\qquad
f_{\mathsf D}=0.082682.
\]
The fractured fraction reaches its maximum
\[
f_{\mathsf F}^{\max}=0.572917
\]
at the sampled value \(\Gamma_0=0.094737\).

Panel (d) shows the representative cut at \(s=0.25\). The maximum concurrence can remain above \(C_{\min}\) where \(\mathcal O_{\max}\leq0\), demonstrating that entanglement survival does not imply simultaneous Bell violation, teleportation advantage, and positive coherent information.

This result is important because it separates two statements that are sometimes treated as equivalent. A gravitational cat state may remain entangled and yet fail as a Bell-nonlocal, teleportation-capable, and transmission-useful channel. In this sense the finite-width gravcat channel exhibits resource fracture: entanglement can survive beyond the region where the stronger operational channel conditions are satisfied.

\section{Conclusion}
\label{sec:Conclusion}
We have studied a finite-width gravitational cat-state channel and asked when gravity-mediated entanglement remains useful in an operational sense. Instead of using entanglement alone as the final benchmark, we tested the same generated state through Bell violation, teleportation fidelity, coherent information, Fisher information, and state distinguishability. This gives a stricter view of usefulness, because an entangled state may still fail to provide a clear communication, metrological, or discrimination advantage.
The results show that finite wave-packet width and spatially correlated dephasing strongly affect the useful operating window. In the scaled regime considered here, there are finite time intervals where the gravitationally generated state is both Bell-nonlocal and teleportation-useful. However, these intervals become smaller when the wave packets are broader or when dephasing is stronger. Thus, finite width should not be treated only as a small correction to the ideal point-particle model; it directly changes the conditions under which the gravitational channel remains useful.

We also showed that residual electrostatic interactions can be especially important. They do not merely add random noise. Because they can contribute coherently to the same nonlocal branch phase as gravity, they may reduce or even cancel the gravitational signal. The resulting cancellation band gives a practical tolerance condition for future gravity-mediated-entanglement tests: electrostatic effects must be controlled not only in strength, but also in their coherent phase contribution.

Finally, the resource-fracture analysis shows that entanglement survival is not enough to certify the strict same-time operational condition. There are regimes where concurrence remains above the numerical tolerance while at least one of Bell violation, teleportation advantage, or positive coherent information fails. State distinguishability is analyzed separately and is not part of the \(\mathsf U/\mathsf F/\mathsf D\) classification. The present work should therefore be read as a channel-level study of finite-width and residual-force limitations, not as a complete experimental feasibility proposal. Its main message is that realistic tests of gravity-mediated entanglement must account for finite wave-packet width, correlated dephasing, and coherent electrostatic systematics when judging whether the observed state is operationally useful and distinguishable from the benchmark models considered here.

\section*{Acknowledgment}
Jia-Xin Peng is  supported by the National Natural Science Foundation of China (Grant No.~12504566), the Basic Research Program of Jiangsu (Grant No.~BK20250947), the Natural Science Foundation of the Jiangsu Higher Education Institutions (Grant No.~25KJB140013)  and the Natural Science Foundation of Nantong City (Grant No.~JC2024045).

\appendix
\section{Numerical implementation and operational diagnostics}
\label{app:numerics}

This appendix records the numerical procedure used to generate the maps in the main text. It does not redefine the physical model. The finite-width kernel, branch rates, input state, dephasing matrix, operational quantities, and discrimination measures are defined once in Sec.~\ref{sec:framework}. Here we specify how those definitions are assembled into finite-dimensional arrays, optimized over the stated parameter windows, and checked for numerical stability.

\subsection{Branch ordering and rate construction}

The computational basis is ordered as \((LL,LR,RL,RR)\). In the notebook, the variable named \texttt{L} is the geometric branch separation \(\Delta x\), not the logical branch label. The coordinates and resulting distances are those in Eqs.~\eqref{eq:framework_branch_coordinates} and \eqref{eq:framework_branch_distances}. For every width, the code constructs
\begin{equation}
    \begin{split}
   \boldsymbol{\omega}^{G}
    &=
    \bigl(
    \omega_{LL}^{G},
    \omega_{LR}^{G},
    \omega_{RL}^{G},
    \omega_{RR}^{G}
    \bigr)^{\mathsf T},
    \\
    \boldsymbol{\omega}^{E}
    &=
    \bigl(
    \omega_{LL}^{E},
    \omega_{LR}^{E},
    \omega_{RL}^{E},
    \omega_{RR}^{E}
    \bigr)^{\mathsf T}.
    \label{eq:app_rate_vectors}     
    \end{split}
\end{equation}
The pure-gravity calculation uses
\(\boldsymbol{\omega}=\boldsymbol{\omega}^{G}\), the contamination calculation uses
\(\boldsymbol{\omega}=\boldsymbol{\omega}^{G}+\epsilon\boldsymbol{\omega}^{E}\), and the metrological calculation uses
\(\boldsymbol{\omega}(g)=g\boldsymbol{\omega}^{G}+\epsilon\boldsymbol{\omega}^{E}\).
The invariant nonlocal rate is evaluated by
$(1,-1,-1,1)\cdot\boldsymbol{\omega}$.
The mean-field benchmark is obtained by arranging
\(\boldsymbol{\omega}^{G}\) into \(W_G\) and applying
Eq.~\eqref{eq:local_projection_matrix}. The stochastic benchmark is evaluated independently with
Eqs.~\eqref{eq:framework_stochastic_state} and
\eqref{eq:framework_Qst}; it does not receive an additional factor
\(\exp[-\Gamma_0Q_{ij}\tau]\). Its Bell and teleportation curves are the separate pointwise envelopes in
Eq.~\eqref{eq:framework_stochastic_envelopes}.

\subsection{Elementwise channel evolution}

The initial matrix is the rank-one projector in Eq.~\eqref{eq:framework_initial_state}. For the gravitational, mean-field, electrostatic-contamination, metrological, and resource-fracture calculations, the \(4\times4\) dephasing array is constructed from Eqs.~\eqref{eq:framework_Qij} and \eqref{eq:framework_branch_displacements}. Define the rate-difference matrix
\begin{equation}
    \Delta W
    =
    \boldsymbol{\omega}\,\mathbf 1^{\mathsf T}
    -
    \mathbf 1\,\boldsymbol{\omega}^{\mathsf T},
    \qquad
    \mathbf 1=(1,1,1,1)^{\mathsf T}.
    \label{eq:app_rate_difference}
\end{equation}
The state is evaluated directly as
\begin{equation}
    \rho(\tau)
    =
    \rho(0)
    \odot
    \exp[-i\tau\Delta W]
    \odot
    \exp[-\Gamma_0\tau Q],
    \label{eq:app_elementwise_evolution}
\end{equation}
where \(Q=(Q_{ij})\) denotes the \(4\times4\) dephasing matrix defined by Eq.~\eqref{eq:framework_Qij}, and \(\odot\) denotes elementwise multiplication. This is the array implementation of Eq.~\eqref{eq:framework_branch_evolution}, not a second evolution law.

\begin{table*}[t]
\scriptsize
\caption{Numerical grids used to generate the seven figures.}
\label{tab:numerical_grids}
\begin{ruledtabular}
\begin{tabular}{lcccl}
Figure & \(N_s\) and \(s\)-range & \(N_\tau\) and \(\tau\)-range & Additional grid & Fixed notes\\
\hline
\ref{fig:fig1}(a--c) & \(110,\ [0.035,0.55]\) & \(150,\ [0,18]\) & --- & \(\Gamma_0=0.018\)\\
\ref{fig:fig1}(d) & \(105,\ [0.04,0.55]\) & \(260,\ [0.60,18]\) & \(N_\Gamma=120,\ [0,0.22]\) & ---\\
\ref{fig:fig2} & \(105,\ [0.04,0.55]\) & \(260\) & \(N_\Gamma=115,\ N_\eta=41\) & \(\eta\in[-0.98,0.98]\)\\
\ref{fig:fig3} & \(90,\ [0.04,0.55]\) & \(210,\ [0.60,18]\) & \(N_\epsilon=260,\ [-1.20,5.20]\) & ---\\
\ref{fig:fig4} & \(72,\ [0.04,0.55]\) & \(180,\ [0.60,18]\) & \(N_\epsilon=200,\ [-1.20,4.766]\) & ---\\
\ref{fig:fig5} & \(46,\ [0.04,0.55]\) & \(120,\ [0.60,18]\) & \(N_\Gamma=46,\ N_\epsilon=88\) & \(\Gamma_0\in[0,0.16]\)\\
\ref{fig:fig6} & \(80,\ [0.05,0.55]\) & \(140,\ [0.60,18]\) & \(N_q=41,\ [0,1]\) & \(s_{\rm cal}=0.25\)\\
\ref{fig:resource_fracture}(b) & \(50,\ [0.05,0.55]\) & \(130,\ [0.60,18]\) & \(N_\epsilon=96\) & ---\\
\ref{fig:resource_fracture}(c) & \(32,\ [0.05,0.55]\) & \(130,\ [0.60,18]\) & \(N_\Gamma=20,\ N_\epsilon=48\) & \(\Gamma_0\in[0,0.12]\)
\end{tabular}
\end{ruledtabular}
\end{table*}

Floating-point asymmetry is removed through
\begin{equation}
    \rho
    \leftarrow
    \frac{\rho+\rho^\dagger}{2}.
    \label{eq:app_hermitize}
\end{equation}
The analytic construction preserves unit trace in Figs.~{\ref{fig:fig1}--\ref{fig:fig5}. The quantum-Fisher-information (QFI) single-state routine and the discrimination and resource-fracture routines additionally normalize by the computed trace after hermitization. The code does not apply one universal trace-restoration step after every state evaluation.

For the gravitational-amplitude calculation,
Eq.~\eqref{eq:fig4_drho} is evaluated analytically. This avoids a finite-difference step in both the quantum and Pauli Fisher informations.

\subsection{Operational diagnostics and finite-time optimization}

For every state, the nine Pauli correlations are computed directly. The singular values of the \(3\times3\) correlation matrix determine both the Horodecki CHSH value and the teleportation fidelity. Concurrence and coherent information are calculated from the corresponding \(4\times4\) and reduced \(2\times2\) matrices.

The interval \(0.60\leq\tau\leq18\) is used whenever a finite-time operational quantity is optimized. Panels displaying the complete initial evolution use \(0\leq\tau\leq18\), as specified below. All reported maxima, phase boundaries, and copy-complexity values are therefore conditional on these finite scan windows. NumPy's first maximizing index is retained when the sampled maximum is degenerate. The useful duration is evaluated through Eq.~\eqref{eq:Tuse_discrete}. Figure~\ref{fig:fig5} uses the fixed threshold in Eq.~\eqref{eq:Tmin_value}, and Fig.~\ref{fig:resource_fracture} uses \(C_{\min}=10^{-4}\).
The grids implemented in the notebook are:
In Table~\ref{tab:numerical_grids}, each \(N_x\) denotes the number of sampled grid points of the variable indicated by its subscript; thus \(N_s\), \(N_\tau\), \(N_\Gamma\), \(N_\eta\), \(N_\epsilon\), and \(N_q\) are the numbers of sampled points in \(s\), \(\tau\), \(\Gamma_0\), \(\eta\), \(\epsilon\), and the Chernoff minimization parameter \(q\), respectively.
For Fig.~\ref{fig:fig2}, panels (a) and (b) use \(0\leq\tau\leq18\), whereas panels (c) and (d) use \(0.60\leq\tau\leq18\). Figure~\ref{fig:fig5} stores \(\Gamma_c\) only when the accepted set is nonempty. Its robust fraction is evaluated over the finite-\(\Gamma_c\) set in Eq.~\eqref{eq:robust_valid_set}. Figure~\ref{fig:resource_fracture}(c) uses every second point of the 96-point electrostatic grid.

\subsection{Calibrated channel discrimination}

For the calibrated discrimination test, the raw electrostatic rate is evaluated at \(s_{\rm cal}=0.25\), and the single factor \(\alpha_E\) is fixed through Eq.~\eqref{eq:fig6_alpha_calibration}. It is then reused at every test width; no width-dependent re-fitting is performed. The canonical vectors in Eq.~\eqref{eq:fig6_canonical_rates} remove global and local phase offsets before the states are compared.

The trace distance is calculated from the eigenvalues of the Hermitian state difference. For the Chernoff exponent, matrix powers are evaluated spectrally. If
\(\rho=V\operatorname{diag}(\lambda_k)V^\dagger\), the code uses
\begin{equation}
    \rho^q
    =
    V\operatorname{diag}(\lambda_k^q)V^\dagger,
    \qquad 0\leq q\leq1,
    \label{eq:app_spectral_power}
\end{equation}
after setting eigenvalues below
$\lambda_{\rm cut}=10^{-14}$
to zero for the matrix-power operation. The minimization is performed only on the uniform 41-point grid \(q\in[0,1]\); no local refinement is used. The exponent is then maximized over \(0.60\leq\tau\leq18\). The reported \(n_{99}^{\rm est}\) is the real-valued estimate in Eq.~\eqref{eq:fig6_required_copies}; values shown in the plot are capped at \(10^8\).

\subsection{Numerical stability and reproducibility}
\label{app:numerical_stability}

All figures are generated by direct evaluation of \(4\times4\) matrices on the grids listed in Table~\ref{tab:numerical_grids}. Hermitian symmetrization removes roundoff asymmetry. Routines that require spectra replace small negative numerical eigenvalues by zero; the entropy routine clips eigenvalues to \([0,1]\), while the Chernoff matrix-power routine uses the cutoff \(10^{-14}\). These operations are numerical regularizations and are not presented as an independent complete-positivity test.
For the Pauli Fisher information, correlations are clipped to
$-1+10^{-10}
\leq T_{mn}\leq
1-10^{-10}$,
and the denominator is bounded below by \(10^{-12}\). For the QFI spectral sum, terms with \(\lambda_\mu+\lambda_\nu\leq10^{-12}\) are omitted. The code uses the analytic derivative in Eq.~\eqref{eq:fig4_drho}.

The notebook does not perform local refinement of the Chernoff parameter, automatic grid-convergence extrapolation, or universal rejection of states based on a separate positivity test. Accordingly, the phase boundaries, scaling fit, and copy-complexity values should be interpreted as results resolved on the reported grids. Large \(n_{99}^{\rm est}\) values identify a high-copy-cost region rather than precise experimental run counts. A device-level analysis would additionally require calibration uncertainty, readout errors, finite detection efficiency, and uncertainty in the branch geometry.

\bibliographystyle{apsrev4-2}
\bibliography{ref}

\end{document}